%% file: main.tex
\documentclass{article}
\usepackage[utf8]{inputenc}
\usepackage{lscape}
\usepackage[margin=1.0in]{geometry} % showframe
\usepackage{graphicx}
\usepackage{booktabs}
\usepackage{longtable}
\usepackage{tabularx}
\usepackage{amsmath}
\usepackage[colorlinks=true,urlcolor=purple,
linkcolor=purple,citecolor=purple]{hyperref}
\usepackage{multicol}
\usepackage[skip=0pt, labelfont=bf, font=footnotesize]{caption}
\usepackage{threeparttable}
\usepackage{threeparttablex}
\usepackage[capposition=bottom]{floatrow}
\usepackage[T1]{fontenc}
\usepackage{etoolbox}
\BeforeBeginEnvironment{tabular}{\small}
\usepackage{nameref}
\usepackage{tikz}
\usepackage[bottom]{footmisc}
\usepackage{authblk}
\usepackage[super,sort&compress]{natbib}
\usepackage{float}
\floatstyle{plaintop}
\restylefloat{table}

\makeatletter
\@fpsep\textheight
\makeatother

\title{Conceptual structure and the growth of scientific knowledge\thanks{E-mail \href{mailto:gkedri001@umn.edu}{kedri001@umn.edu} or \href{mailto:rfunk@umn.edu}{rfunk@umn.edu}. }}
\author[1]{Kara Kedrick}
\author[2]{Ekaterina Levitskaya}
\author[3]{Russell J. Funk}
\affil[1]{Institute for Complex Social Dynamics, Carnegie Mellon University}
\affil[2]{Wagner Graduate School of Public Service, New York University}
\affil[3]{Carlson School of Management, University of Minnesota}
\date{}

\begin{document}

\maketitle

\begin{abstract}
How does scientific knowledge grow? This question has occupied a central place in the philosophy of science, stimulating heated debates, but yielding no clear consensus. Many explanations can be understood in terms of whether and how they view the expansion of knowledge as proceeding through the accretion of scientific concepts into larger conceptual structures. Here, we examine these views empirically, performing a large-scale analysis of the physical and social sciences, spanning five decades. Using natural language processing techniques, we create semantic networks of concepts, wherein noun phrases become linked when used in the same paper abstract. For both the physical and social sciences, we observe increasingly rigid conceptual cores (i.e., densely connected sets of highly central nodes) accompanied by the proliferation of periphery concepts (i.e., sparsely connected nodes that are highly connected to the core). Subsequently, we examine the relationship between conceptual structure and the growth of scientific knowledge, finding that scientific works are more innovative in fields with cores that have higher conceptual churn and with larger cores. Furthermore, scientific consensus is associated with reduced conceptual churn and fewer conceptual cores. Overall, our findings suggest that while the organization of scientific concepts is important for the growth of knowledge, the mechanisms vary across time.

\end{abstract}

\pagebreak

\section{Introduction}

The past century has witnessed an unprecedented expansion of scientific knowledge. As early as the 1960s, for example, pioneering sociologist of science de Solla Price observed a doubling in the growth of journals every 13 years \citep{price1961} and abstracts every 15 \citep{price1963}. Subsequent research suggests that this expansion has continued ever since, and is characteristic not only of journals and abstracts, but also extends to scientific concepts and many other indicators \citep{bornmann2020, milojevic2015quantifying, Tabah1999es}. Yet, while few would dispute the dramatic growth of scientific knowledge, there is far less consensus on the underlying mechanisms. Indeed, the question of how, precisely, scientific knowledge grows has stimulated generations of debate among philosophers, with the literature being characterized by many diverse, seemingly incompatible views \citep{kuhn1962, lakatos_musgrave_1970, laudan1978, popper1963}.  
 
In this study, we contribute to these debates by examining the growth of scientific knowledge empirically. Despite their outward differences, many proposed explanations for the growth of scientific knowledge can be understood in terms of whether and how they view the expansion of knowledge as proceeding through the accretion of scientific concepts into larger conceptual structures. Classical perspectives, for instance, tend to suggest that scientific knowledge grows through the expansion and refinement of a conceptual core \citep{cole1994,cole2006,kuhn1962,popper1963}. Other views see science as more pluralistic, suggesting that the growth of knowledge comes through the proliferation of competing alternatives, manifesting in the form of multiple cores or research programs \citep{Gonzalez2015,lakatos_musgrave_1970, laudan1978}. Furthermore, some theorize that scientific advancement is marked by a fixed set of core concepts \citep{ lakatos_musgrave_1970}, while others suggest that churn or change of core concepts is the hallmark of the growth of knowledge \citep{chu_2021,laudan1978,popper1963}. In short, while scholars differ sharply in their views on the underlying mechanisms of scientific growth, there is general agreement that growth is likely to manifest in the conceptual structure of science. Therefore, empirical analysis of the conceptual structure of science may add a novel lens on the growth of knowledge, thereby enabling future philosophical advance.
 
\begin{figure}[ht!]
    \centering
    \includegraphics[width=14cm]{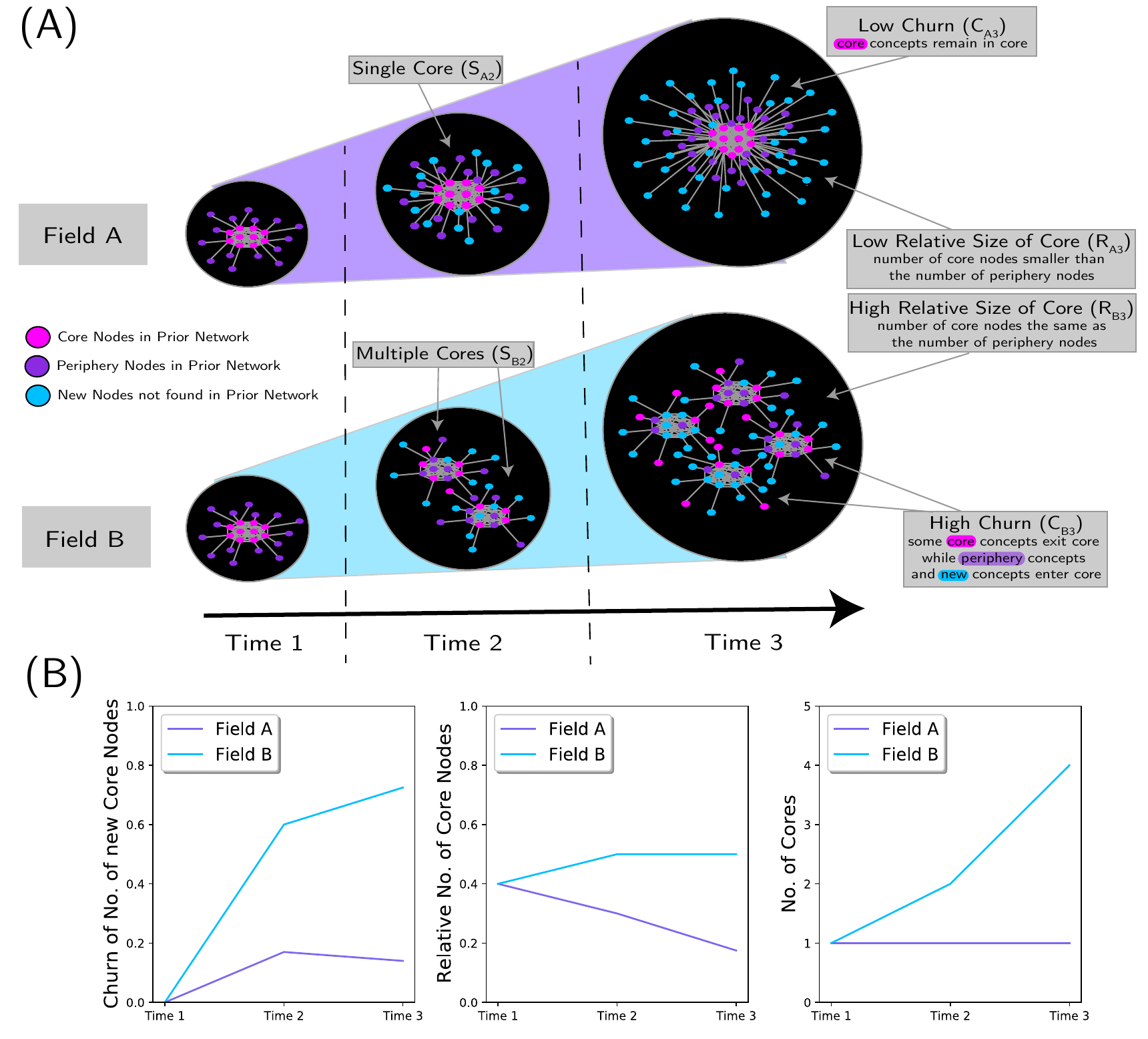}
    \caption{\textbf{Schematic illustration of core/periphery measures over time for two fields. (1A)} illustrates different ways in which core/periphery structures can unfold across time. The highly connected nodes at the center of each network represent the core, whereas the relatively-less connected nodes toward the edge of the networks represent the periphery. The color of the nodes indicates the nodes' status from the prior year---pink indicates that the node was in the core, purple indicates that the node was in the periphery, and blue indicates that the node was added to the network at time t. At t=1, Field A and Field B are identical. As time progresses, each field differs in terms of our core/periphery measures. Field A maintains a single core across time, while additional cores emerge in Field B (i.e., there are two cores at t=2 and four cores at t=3). The fields also differ in terms of the churn in core nodes. Field A has low churn; the pink-core nodes remain in the core and a small number of purple-periphery nodes and blue-new nodes are added to the core. In contrast, the core nodes in Field B change over time; that is, purple nodes—which were in the periphery during the prior year—and blue nodes—which are new to the network—enter the core. Finally, the fields differ in terms of the relative size of the core. Field A has a relatively smaller core, driven by growth in the periphery. Field B has a relatively larger core, with similar numbers of core and periphery nodes. \textbf{(1B)} plots how the fields change in terms of the core/periphery measures across time. 
 }
\label{fig:schematic_illustration}
\end{figure}

To that end, we conducted a large-scale analysis of 582,679 physical sciences articles from the American Physical Society (1965-2016) and 2,022,545 social sciences articles from the Web of Science (1993-2016). For each year and disciplinary subfield, we created a semantic network of concepts, wherein concepts were linked if they appeared in the same abstract. Edge weights were determined by the frequency in which pairs of concepts co-occurred within the same abstract \citep{Newman_2001}. Motivated by insights from the philosophy of science \citep{cole1994,kuhn1962,popper1963,lakatos_musgrave_1970,laudan1978}, we studied these networks in terms of their core/periphery organization, wherein core nodes consist of a densely connected set of highly central nodes and periphery nodes consist of sparsely connected nodes that are highly connected to the core. We characterized core/periphery organization according to three parameters (see \emph{Materials and Methods} and Fig. \ref{fig:schematic_illustration} for more information). The first parameter, \(C_{it}\), captured the churn or change of core nodes in a network. Specifically, \(C_{it}\) addressed the opposing positions voiced by philosophers in regard to the churn of core concepts, with some theorizing that a fixed set of core concepts is beneficial to scientific fields \citep{lakatos_musgrave_1970, latour1987}, and others suggesting turnover is advantageous and drives the growth of knowledge \citep{chu_2021,laudan1978}. The second parameter, \(R_{it}\), measured the relative size of the cores, and is defined as the fraction of concepts that comprise the cores out of the total number of concepts in a network. \(R_{it}\) addresses the theory that a set of general core concepts may promote the creation of more specific conceptual material (e.g., theoretical extensions, auxiliary hypotheses, and empirical applications), resulting in the expansion of the periphery and decline in the relative size of the core \citep{lakatos1978}. The third parameter, \(S_{it}\), quantifies the number of core/periphery structures in a network. \(S_{it}\) addresses debates in philosophy of science about the number of cores, with some authors depicting scientific fields as having a single set of core concepts \citep{cole1994,cole2006,kuhn1962}, and others suggesting that multiple cores are more common \citep{Gonzalez2015,lakatos_musgrave_1970,laudan1978}.

To detect core/periphery structures, we used an algorithm that was capable of identifying multiple cores \citep{kojaku2017}. The algorithm is based on Borgatti and Everett's (\citeyear{borgatti2000}) notion of the idealized core, wherein core nodes are maximally connected to both other core nodes and periphery nodes, with minimal interconnectivity among nodes in the periphery. Subsequently, we assess the dynamics of these core/periphery measures across time and scientific fields, according to the three properties. 

After documenting changes in the core/periphery organization of scientific concepts, we then consider the relationship between the three core/periphery properties and two key processes in the growth of scientific knowledge: innovation and consensus. Innovation is fundamental to the growth of scientific knowledge, enabling increases in both the depth and breadth of understanding \citep{funk2017}. Scientific consensus has also been widely viewed as an indicator of growth, through which scientific communities unify distinct or competing perspectives \citep{mulkay1975,wimsatt2006}. Our findings suggest that innovation and scientific consensus are associated with fluctuations in the core/periphery properties.

\section{Results}

\subsection{Properties of Scientific Concepts}\label{sec:properties}
To better understand the nature of the nodes in our concept networks, we begin our analysis by conducting a series of investigations comparing the properties of scientific concepts that appear in the core versus the periphery. Consistent with intuition, we found that core concepts are generally more established within a scientific field than periphery concepts (S2.1 Appendix). Specifically, we found that core concepts from the social sciences appeared in more subfields (Welch's \emph{t} = 28.12, \emph{p} < 0.001) and were used in more papers (Welch's \emph{t} = 6.11, \emph{p} < 0.001). For the physical sciences, we found that core concepts were more likely than periphery concepts to be older (Welch's \emph{t} = 5.56, \emph{p} < 0.001) and used in more papers (Welch's \emph{t} = 18.42, \emph{p} < 0.001). Core concepts were also more likely to appear among the 892 curated topical tags on the Physics Stack Exchange Q\&A site, indicating their widespread use even beyond the community of academic physics research (Welch's \emph{t} = 12.54; \emph{p} < 0.001). We also find that periphery concepts have a higher level of specificity when compared to core concepts. For the social sciences, periphery concepts have a greater number of characters (Welch's \emph{t} = -19.70, \emph{p} < 0.001) and words (Welch's \emph{t} = -15.43, \emph{p} < 0.001; S2.1 Appendix). For the physical sciences, periphery concepts have a greater number of characters (Welch's \emph{t} = -9.85, \emph{p} < 0.001; S2.1 Appendix) and digits (Welch's \emph{t} = -8.78, \emph{p} < 0.001; S2.1 Appendix). Our findings are in line with Lakatos's (\citeyear{lakatos1978}) characterization of periphery concepts as being relatively more specific than core principles. This increased specificity allows for the abstract/theoretical core to be empirically tested via periphery concepts or auxiliary hypotheses. In short, core and periphery concepts have distinctive properties: core concepts tend to be more established and abstract in nature (e.g., ``vaccination''), whereas periphery concepts are relatively less established and have greater specificity (e.g., ``diastolic blood pressure'').  

\subsection{Conceptual Structure Over Time}\label{sec:over_time}
As noted previously, while philosophers of science suggest that the organization of scientific concepts is integral to the growth of scientific knowledge, they differ sharply on the details. In this section, we consider the possibility that the organization of scientific concepts may change over time, as scientific theories evolve, and or across disciplines, in which case, disagreements from prior work may be reconcilable. To simplify our presentation, we examined how the three parameters outlined above---(1) the churn of core nodes, (2) the relative size of the cores, and (3) the number of core nodes---changed by averaging measures across subfields. We excluded subfields that had fewer than 100 concepts over the entire time period of analyses.

\begin{figure}[ht!]
    \centering
    \includegraphics[width=14cm]{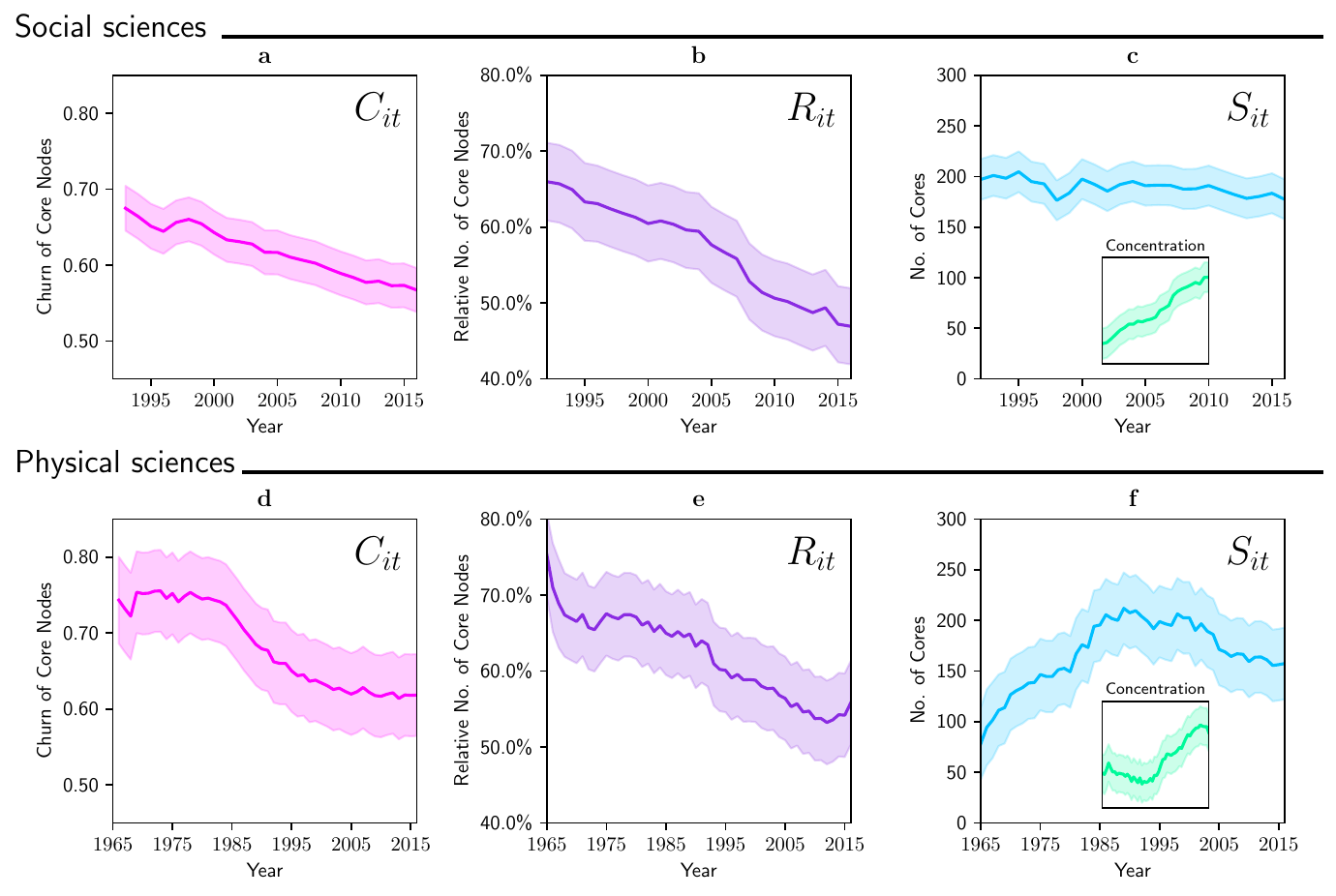}
    \caption{\textbf{Conceptual structures over time.} The plots show changes in the core/periphery organization averaged across subfields. For both the social sciences (top) and physical sciences (bottom), the plots show a decrease in the amount of churn of core concepts (a,d) and a decrease in the relative number of core nodes (b,e). However, the social and physical sciences differ in terms of the number of cores over time. We see a slight decrease in the number of cores for the social sciences (c). Alternatively, we see an overall increase in the number of cores for the physical sciences (f). The inset plots (found in plots c and f) show that concepts become increasingly concentrated in a few number of core/periphery structures across time. 
 }
\label{fig:OverTime}
\end{figure}

\begin{landscape}
\begin{figure}[ht!]
    % \centering
    \includegraphics[width=20cm]{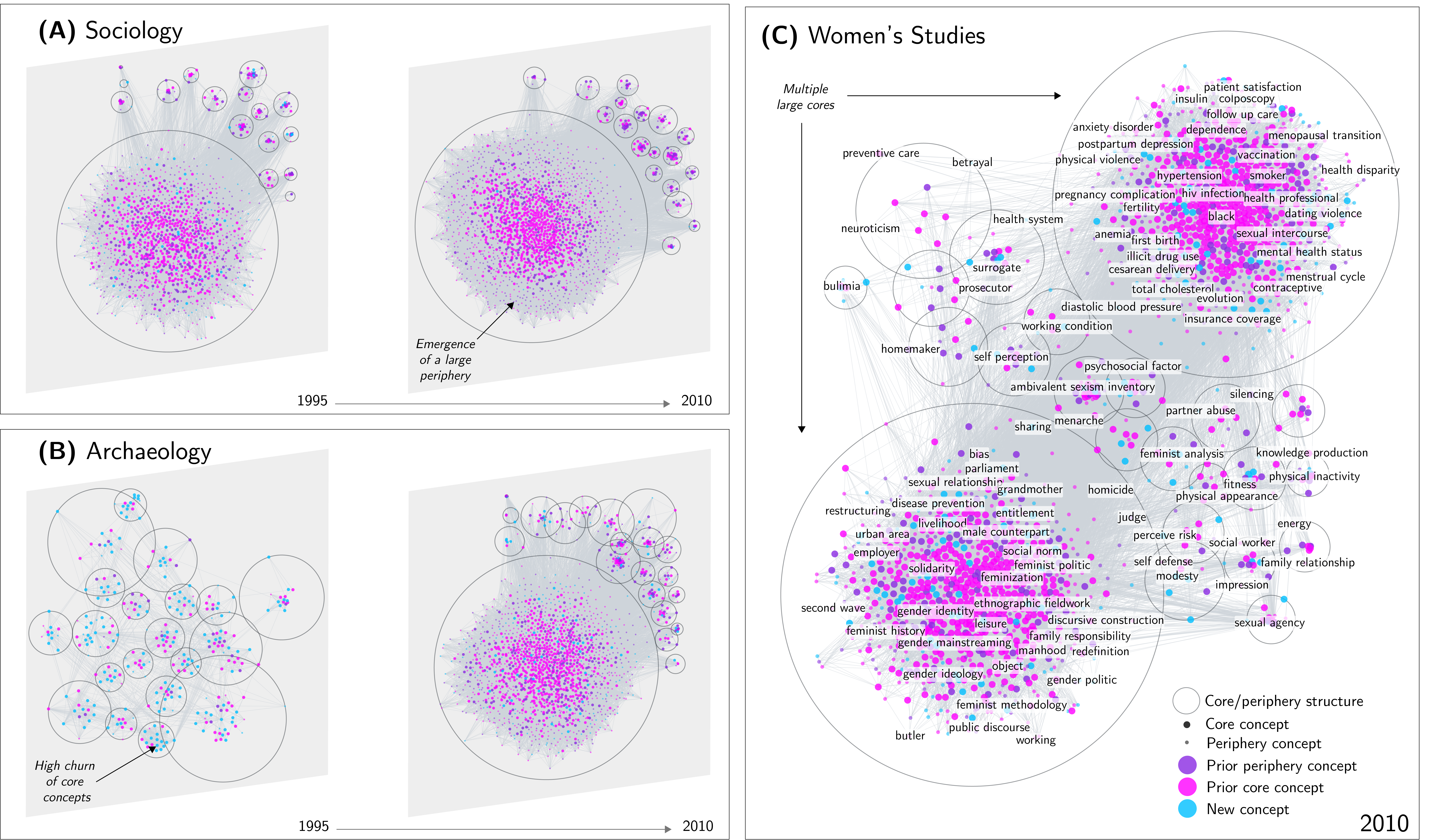}
    \caption{\textbf{Schematic illustration of core/periphery measures.}  \textbf{(1A)} Comparing the conceptual structure of Sociology in 1995 and 2010 reveals that the relative size of the core decreases over time. Specifically, there are more periphery nodes in 2010 (visually indicated as smaller in size) causing the relative size of the core to shrink. \textbf{(1B)} Comparing the structure of Archaeology in 1995 and 2010 reveals a decrease in the churn of core concepts. Core nodes in the 1995 network include many new concepts (blue) and prior periphery concepts (purple) in addition to prior core concepts (pink). In contrast, the core in the 2010 network is composed primarily of nodes that were previously in the core (pink). \textbf{(1C)} The conceptual structure of Women's Studies 2010 provides an example of a multi-core network, wherein the majority of concepts are concentrated around two large core/periphery structures, one corresponding to more humanistic approaches (bottom left), and the other more biomedical ones (top right). \emph{Note:} Due to the computational burden of generating large-scale network diagrams, we could not incorporate all the concepts present in our analytical data into this visual representation.
 }
\label{fig:schematic_social}
\end{figure}
\end{landscape}

 First, we found a decline in churn (i.e., change in core concepts), \(C_{t}\) (Fig.~\ref{fig:OverTime}). We verify the decline in churn over time through regression models, which show that year has a statistically significant and negative relationship with churn of core concepts for the social and physical sciences (\(p < 0.001\); Table 1 Appendix). To quantify the decline, we calculated the percent change in churn for the social sciences (i.e., \([(C_{1993} - C_{2016})/C_{1993}]\times100\)) and for the physical sciences (i.e., \([(C_{1966} - C_{2016})/C_{1966}]\times100\)). We found that the degree of churn of core concepts decreased on average by \(15.96\%\) for the social sciences and by \(20.03\%\) for the physical sciences. These findings support the claim that core concepts change over time (\citealp{laudan1978}; illustrated in Fig. \ref{fig:schematic_social}). However, the amount of change decreases as scientific fields develop, suggesting that the core composition becomes increasingly rigid.

Next, we found a decline in the relative number of core concepts, \(R_{t}\) (Fig.~\ref{fig:OverTime}). We verify the decline in relative size of the cores over time through regression models, which show that year has a statistically significant and negative relationship with the relative number of core concepts for the social and physical sciences (\(p < 0.001\); Table 1 Appendix). To quantify the decline, we calculated the percent change in the relative size of the cores for the social sciences (i.e.,  \([(R_{1992} - R_{2016})/R_{1992}]\times100\)) and for the physical sciences (\([(R_{1965} - R_{2016})/R_{1965}]\times100\)). We found that the relative number of core nodes decreased by \(28.86\%\) for the social sciences and by  \(25.93\%\) for the physical sciences. These results suggest that over time, new periphery concepts are added, causing the relative number of core nodes (with respect to the total number of concepts in the network) to decline (illustrated in Fig. \ref{fig:schematic_social}). This ``shrinking'' of the conceptual cores may happen because the rigidity of the core prevents new concepts from entering, leaving them instead to end up in the periphery. Another possibility is that scientists are introducing various ways to empirically test core concepts through the advent of specific conceptual material (e.g., auxiliary hypotheses; \citeauthor{lakatos1978}, \citeyear{lakatos1978}), causing growth of scientific concepts to occur through the expansion of the periphery. 

Finally, our algorithm detected multiple core/periphery structures, \(S_{it}\), for each subfield \(i\) at time \(t\); however, the social and physical sciences differed in terms of how the number of cores changed over time (Fig.~\ref{fig:OverTime}). Specifically, we see a slight decline in the number of cores for the social sciences and an overall increase in the number of cores for the physical sciences (with a leveling off and modest decline beginning in the mid-1990s, around the time when our data for the social sciences begin). We verify these trends over time through regression models, which show that year has a statistically significant and negative relationship with number of cores for the social sciences (\(p < 0.001\)) and positive relationship for the physical sciences (\(p < 0.001\); Table 1 Appendix). To quantify the decline for the social sciences, we calculated the percent change (i.e., \([(S_{1992} - S_{m,2016})/S_{1992}]\times100\)), and found that the social sciences had 19.73 fewer cores (with a percent change of \(-10.01\%\)). Although the observed trend indicates a discernible and statistically significant decrease, the magnitude of this decline appears to be relatively minor. Furthermore, our analysis reveals that this pattern is not consistent across various networks, including those defined by different thresholds, nor is it consistent when employing an alternative core/periphery detection algorithm (refer to S2.2 in the Appendix). Consequently, asserting a temporal trend in the fluctuation of the number of cores within the social sciences is challenging. To quantify the increase for the physical sciences, we calculated the percent change (i.e., \([(S_{2016} - S_{1992})/S_{1992}]\times100\)), and found that there were 79.15 more cores with a percent change of \(101.18\%\). 

To better understand these trends, we examined how changes in the number of cores compared to the concentration of concepts within particular core/periphery structures over time. We used a Herfindahl index as a measure of concentration, wherein higher values indicate that concepts are concentrated around a single or small number of core/periphery structures. As shown in the inset plot of Fig.~\ref{fig:OverTime}, concentration is generally increasing for the social sciences (with an average percent change of 227.95\%) and the physical sciences (with an average percent change of 214.62\%), which suggests that over time concepts are assembling into a smaller number of larger or dominant core/periphery structures. Thus, although the number of core/periphery structures in the physical sciences is increasing over time overall, the newly added cores are relatively small. Alternatively, we see a decline in the number of cores for the social sciences, which suggests that as concepts become increasingly concentrated around larger core/periphery structures, smaller core/periphery structures may be eliminated from the network.

\subsection{Conceptual Structure and Scientific Innovation}\label{sec:innovation}
Our analyses thus far have documented changes in the organization of scientific concepts over time. In this section, we examine the implications of those changes for the growth of knowledge. Innovation---a key mechanism underpinning the growth of scientific knowledge---involves the advent of new ideas that reshape the theoretical and/or methodological landscape of a scientific field. Yet, little empirical research has examined whether and how the structural properties of scientific concepts foster innovation. Our work investigates whether the properties of a field's conceptual core can foster discoveries that push the scientific frontier.

To quantify innovation, we used a recently developed measure called the CD index \citep{funk2017}, which captures the degree to which a paper pushes the scientific frontier by disrupting the streams of knowledge on which it builds. More specifically, a scientific paper is considered to be highly disruptive if the papers that cite it are less likely to cite its references. Disruptiveness was calculated for all papers as of 2017 (indicated by \(CD_{2017y}\)). The \(CD_{2017y}\) index is in line with expert assessments of disruptive work and has been thoroughly validated by prior research \citep{bornmann2020, funk2017, wu2019large}. 

We examine the relationship between the core/periphery organization of scientific concepts and innovation using linear regression models, estimated separately for the social sciences (Models 1-2; Table~\ref{table:innovation_table}) and physical sciences (Models 3-4; Table~\ref{table:innovation_table}) with the churn of core nodes \(C_{it}\), the relative size of the cores \(R_{it}\), and the number of cores \(S_{it}\) as our primary independent variables. For all of our models, we included fixed effects for year and subfield. Model 2 and Model 4 included the number of authors and the number of references as additional controls.

First, we examined whether the churn of core concepts is associated with innovation. We found that an increase in the churn of core nodes, \(C_{it}\), was positively associated with the \(CD_{2017y}\) index, for both the social sciences (\(\textrm{Model 1:} \beta = 0.2939, SE = 0.0135, CI = [0.2674,0.3204], p < 0.001; \textrm{Model 2:} \beta = 0.3335, SE = 0.0132, CI = [0.3076,0.3594], p < 0.001\); Table \ref{table:innovation_table}) and the physical sciences (\(\textrm{Model 3:} \beta = 0.1697, SE = 0.0189, CI = [0.1326,0.2068], p < 0.001; \textrm{Model 4:} \beta = 0.1729, SE = 0.0191, CI = [0.1353,0.2104], p < 0.001\)). High churn or change may indicate that a field is less committed to a core set of concepts, and  therefore existing works are more vulnerable to displacement by newer works. Alternatively, churn may stimulate creative thinking, and allow researchers to develop ideas that disrupt a scientific domain.

Second, we considered the association between innovation and the relative size of the cores. We found that an increase in the relative core size (\(R_{it}\); i.e., or comparatively larger cores) was positively associated with the \(CD_{2017y}\) index for both the social sciences (\(\textrm{Model 1:} \beta = 0.0040, SE = 0.0001, CI = [0.0039,0.0042], p < 0.001; \textrm{Model 2:} \beta = 0.0047, SE = 0.0001, CI = [0.0046,0.0049], p < 0.001\); Table \ref{table:innovation_table}) and the physical sciences (\(\textrm{Model 3:} \beta = 0.0006, SE = 0.0001, CI = [0.0003,0.0008], p < 0.001; \textrm{Model 4:} \beta = 0.0007, SE = 0.0001, CI = [0.0004,0.0009], p < 0.001\)). These results suggest that a field is ripe for innovation when the set of core concepts is relatively large, perhaps indicating that core concepts have not been significantly developed (e.g., through theoretical extensions, auxiliary hypotheses, and empirical applications, which may result in a larger periphery\citep{lakatos1978}). That is, there is less evidence supporting the core principles, making them subject to change.

Third, we tested whether the number of core/periphery structures within a network influences innovation. We found that the \(CD_{2017y}\) index was associated with an increase in the number of core/periphery structures, \(S_{it}\), for the social sciences (\(\textrm{Model 1:} \beta = 0.0001, SE = 0.0000, CI = [0.0001,0.0001], p < 0.001;\textrm{ Model 2}: \beta = 0.0002, SE = 0.0000, CI = [0.0001,0.0002], p < 0.001\)). However, the results were insignificant for the physical sciences (\(\textrm{Model 3:} \beta = -0.0000, SE = 0.0000, CI = [-0.0000,0.0000], p = 0.54; \textrm{Model 4:} \beta = -0.0000, SE = 0.0000, CI = [-0.0000,0.0000], p = 0.14\); Table 3). The mixed results suggest that the number of cores is a less consistent predictor of innovation than the other core/periphery properties.

Finally, with respect to the control variables, we find a statistically significant, negative association between the number of authors and the \(CD_{2017y}\) index, which is consistent with prior work suggesting that larger teams tend to produce less disruptive work \citep{wu2019large}.

In summary, some conceptual structures appear more conducive to innovation, and ultimately the growth of knowledge. Our results indicate that prominent and less rigid core/periphery structures may provide an environment ripe for innovative ideas. Specifically, innovative works were associated with concept networks with high churn of core concepts, relatively larger cores, and fewer cores. In conjunction with our findings on over time trends, our results suggests that the innovative potential of a field declines over time: as the core increases in rigidity (high churn) and prevents new concepts from entering the core (decrease in relative size of the core). Our findings also suggest that the number of cores slightly decreases over time for the social sciences, potentially suggesting an environment less conducive to innovation. But, these temporal trends lack consistency across different network variations (refer to S2.2 in the Appendix), and thus should be interpreted with caution. 

\subsection{Conceptual Structure and Scientific Consensus}\label{sec:consensus}
Scientific consensus has been widely viewed as an indicator of scientific growth or development \citep{mulkay1975,wimsatt2006}, and has been associated with structural changes, such as a decrease in the number of communities in citation networks \citep{shwed2010}. We therefore theorized that the conceptual structure of scientific fields would be associated with consensus.

To measure scientific consensus, we derived a dictionary of words that expressed consensus (e.g., ``agree'', ``compatible'', ``support''; S1.5 Appendix), using a combination of hand coding and word embeddings trained specifically on our corpus of scientific text. For example, the following statement from one of the abstracts in our sample uses ``agreement'' to indicate consensus: ``We find the number of fingers observed in our simulations to be in excellent agreement with experimental observations and a linear stability analysis reported recently by Smolka and SeGall (2011)'' \citep{mayo_2013}. We only analyzed physical sciences abstracts because, unlike the WoS data, the APS data uniformly denoted in-text citations using XML tags. The final dictionary included 42 consensus words (S1.6 Appendix). Subsequently, we applied this dictionary to abstracts in the APS corpus, which yielded a count of consensus words per paper.

We ran two linear regression models with our core/periphery measures as independent variables and with controls for publication year, subfield, and counts of words by part-of-speech (i.e., nouns, adjectives, verbs, and adverbs). Model 2 differed from Model 1 by including additional controls for the number of authors and the number of references. We conducted our analysis on abstracts that commented on prior work or the state of the field. Our results reveal that an increase in the incidence of consensus-related words was associated with a decrease in both the churn of core nodes (\(\textrm{Model 1:} \beta = -1.4168, SE = 0.4502, CI = [-2.2992,-0.5345], p = 0.0016; \textrm{Model 2:} \beta = -1.3840, SE = 0.4582, CI = [-2.2821,-0.4860], p = 0.0025\)) and the number of cores (\(\textrm{Model 1:} \beta = -0.0009, SE = 0.0004, CI = [-0.0016,-0.0002], p = 0.0098; \textrm{Model 2:} \beta = -0.0007, SE = 0.0004, CI = [-0.0014,-0.0000], p =0.0441\); Table 2 Appendix).

\section{Discussion}
In developing theories about the growth of scientific knowledge, many philosophers of science have recognized the significance of the accretion of scientific concepts into larger conceptual structures \citep{kuhn1962,lakatos_musgrave_1970,laudan1978}. To shed light on these theoretical perspectives, we conducted a large-scale analysis of the conceptual structures of the social and physical sciences, using techniques from semantic network analysis. Our initial analysis of the nodes in our concept networks suggests that core concepts are relatively more established and abstract in nature, whereas periphery concepts are less established and have higher specificity. We further studied these conceptual structures in terms of their core/periphery organization, which we characterize according to three properties: (1) \(C_{it}\), churn of core nodes, (2) \(R_{it}\), relative size of the cores, and (3) \(S_{it}\), number of cores. Across the social and physical sciences, we found that conceptual cores became more rigid over time, as indicated by a decrease in the churn of core concepts. This increasing rigidity is accompanied by the growth of periphery concepts, as indicated by a decline in the relative size of conceptual cores. 

We also found that while both the social and physical sciences tended to have multiple core/periphery structures, there were differences in how the number of cores changed over time, with prevalence slightly decreasing for social sciences and increasing overall for physical sciences. However, both fields showed an increase in the concentration of concepts around a relatively small number of core/periphery structures. Upon validating our over time results, we discovered inconsistencies in the number of cores for the social sciences across different thresholds (see S2.2). In contrast, we observed a high level of consistency over time for all other core/periphery parameters.

After identifying distinct changes in conceptual structures over time, we focused our analyses on two principal indicators of the growth of scientific knowledge: innovation and consensus. First, we considered the consequences of changes in conceptual organization on innovation. We found that increases in the churn of core concepts (\(C_{it}\)) and increases in relative core size (\(R_{it}\)) were associated with discoveries that push the scientific frontier (as indicated by a measure of ``disruptiveness''; \citealp{funk2017}), which suggests that relatively more prominent and less rigid core/periphery structures may promote the growth of knowledge. The number of cores (\(S_{it}\)) was a less robust predictor of innovation. Specifically, for the social sciences, innovation was associated with an increase in the number of cores; whereas in the physical sciences, there was no evident association between the number of cores and innovation.

Second we considered how conceptual organization relates to scientific consensus by conducting an analysis of articles that commented on prior work or the state of the field (e.g., commentaries, replies, letters to the editor, reviews). Our results reveal that a decrease in the churn of core nodes and a decrease in the number of cores was associated with an increase in the frequency of consensus-related words.  As consensus builds within a scientific field, the decrease in churn of core concepts may reflect a growing agreement among scientists on definitions, methodologies, and interpretations. Consequently, the need to alter core concepts diminishes. 
Moreover, a decrease in the number of cores may indicate that consensus is associated with the consolidation of fields around a unified set of theories or models, resulting in fewer, distinct conceptual frameworks. Overall, our results suggest that conceptual structures in science are highly dynamic, with the relative prominence of core/periphery organization associated with changes in innovation and consensus.

Our research offers an alternative perspective on important trends in the progress of science. Specifically, although in recent decades there has been a proliferation of scientific work, there is growing concern that innovative activity is slowing \citep{jones_2009,gordon_2016,Bhattacharya_2020}. Scientific innovation has often been associated with the recombination of existing ideas or concepts. The introduction of new concepts---fueled by the growth in scientific publication---increases the number of potential recombinations \citep{fink2017,tria2014}. Yet, the rate of innovative works that push the scientific frontier has declined over time \citep{bloom2020ideas, horgan2015, Jones2011}. This discrepancy may potentially be reconciled by considering changes in the structure and dynamics of scientific concepts. Core concepts, by definition, are frequently recombined with fellow core concepts and less-central periphery concepts. As supported by our findings, core concepts tend to be more widely used by scientists within and across fields of science. In fields with more rigid conceptual cores, with low churn or change in core concepts, the same set of concepts will be more often used in combinations. However, if the churn of core concepts is high, then a field is more likely to see recombinations that use a wider variety of concepts. Conceptual structures with less rigid cores that allow for a higher number of unique combinations may therefore facilitate innovation and the growth of knowledge. Our results support this claim by showing that innovation is inversely associated with low churn or a stable set of core concepts, and that the churn of core concepts decreases with time. 

The structural organization of concepts may also influence scientists' ability to think creatively. Specifically, scientists may struggle to develop novel ideas when their field contains rigid cores with a long-standing set of core concepts (low churn) that have been frequently applied or tested (low relative size of core). Such core concepts may be treated as “black boxes” \citep{latour1987}, which are accepted without critical examination. Scientists may also develop fixed approaches to solving problems or struggle to use concepts in different ways. In line with this claim, research suggests that prior experience or familiarity can prevent people from thinking creatively while solving problems \citep{duncker1945,JANSSON1991,maier1931,smith1993}. For example, Duncker (\citeyear{duncker1945}) found that people struggle to solve problems when the solution requires them to use familiar objects in a different way (i.e., use a matchbox as a candle stand instead of a box that holds matches). Similar to this finding, familiarity with and frequent application of rigid core principles may hinder scientists' ability to think creatively or reimagine a scientific domain.

\section{Materials and Methods}\label{Methods}

\subsection{Data}

To map the structure and dynamics of scientific concepts, we drew on two large-scale bibliographic databases: the American Physical Society corpus (``APS data'') and the Web of Science (``WoS data''). While WoS includes bibliographic records for nearly all fields of science and scholarship, we focus our attention on those published in the social sciences (i.e., as classified by WoS; See S1.1 Appendix for discussion of limitations). We create our maps separately for subfields of the physical sciences (via the APS data) and the social sciences (via the WoS data), as prior research suggests different domains of knowledge are likely to exhibit distinctive conceptual organization \citep{cole1995}. For the physical sciences, we assign papers (and their associated concepts) to 10 subfields using Physics and Astronomy Classification Scheme (PACS) based on the first/primary PACS code listed for each paper. For the social sciences, we assign papers (and their associated concepts) to 24 subfields using subject areas from the Web of Science, which correspond to generally recognized academic disciplines (e.g., ``Sociology''). We limit our attention to the period from 1965-2016 and 1992-2016 for the APS and WoS data, respectively. After subsetting and removing articles without abstracts, our analysis is based on a total of 2,605,224 papers, with 582,679 and 2,022,545 coming from the APS and WoS data, respectively. More details on the data are described in the S1.1 Appendix.

\subsection{Mapping Conceptual Structure using Semantic Network Analysis}\label{sec:knowledge_networks}

To identify scientific concepts and map their connections, we used techniques from semantic network analysis, a method of natural language processing. In a semantic network, nodes represent words or concepts; connections (or edges) are recorded among nodes when they co-occur in text (e.g., they appear in the same sentence, abstract, or document). Semantic network analysis has been widely used in prior research to study a variety of phenomena \citep{christianson2020architecture,dworkin2019emergent,Rule2015}. Building on this previous literature, we extracted concepts from the abstracts of scientific articles using the spaCy and Textacy Python packages \citep{honnibal2017, dewilde2021}. For our study, we identified concepts as noun phrases (i.e., nouns together with their dependents). We extracted single-word and multi-word noun phrases, using chunking. We then processed the noun phrases by lemmatizing each token, and removing spaces, stop words, and punctuation (see S1.2 Appendix). Examples of concepts extracted directly from our WoS abstracts include “religious service attendance”, “social capital”, and “youth unemployment.”   

After deriving a final list of noun phrases, we created concept networks by linking two phrases if they appeared in the same abstract. The edges were weighted based on the frequency with which the pair of concepts co-occurred. Self-loops were not permitted. We constructed networks for each subfield: 24 subfields for the WoS data and 10 subfields for the APS data. The networks were remapped every year using a one-year moving window (see Figure 5 Appendix for three-year moving window results).

To reduce noise, semantic network analyses commonly focus on a subset of concepts parsed from the raw text data \citep{Rule2015, hofstra2020}. We follow the approach of prior research and use frequency thresholds to prune extremely common (e.g., ubiquitous scientific phrases) and extremely rare concepts (e.g., typos and parse errors) from our data \citep{hofstra2020}. We began our process by excluding concepts that appeared in only a single article. Then, we generated 25 versions of our networks, wherein concepts were removed based on two thresholds: if they appeared in less than a certain percentage of articles (0.00\%, 0.001\%, 0.025\%, 0.05\%, and 0.1\%) and if they were present in more than a certain percentage of articles (1\%, 5\%, 7.5\%, 10\% and 100\%). For each version, we executed the core/periphery algorithm 10 times. We then averaged each core/periphery measure across these iterations for every version to produce a final average value.

Our analysis centered on the minimum and maximum thresholds that yielded the highest mean $Q^{cp}$-value (see S1.4 Appendix) across 10 iterations, signifying that the thresholds established networks most closely aligned with the ideal core/periphery structure. We also confirmed that, from a qualitative standpoint, the core/periphery structures are well-defined at these thresholds. Specifically, we focused on the version where concepts occurred in more than 0.1 percent and fewer than 1.0 percent of abstracts within a subfield. On average, abstracts from the physical sciences yielded 1,889.30 concepts per subfield, while those from the social sciences produced 2,510.46 concepts per subfield (see Figure 3 Appendix results using our 25 alternative thresholds).

\subsection{Characterizing Conceptual Structures}\label{sec:cp_measures}
We studied conceptual structures in terms of their core/periphery organization, motivated by insights from the philosophy of science \citep{kuhn1962,lakatos_musgrave_1970,laudan1978}. To detect core/periphery structures, we used an algorithm that had the potential to identify multiple cores\citealp{kojaku2017} (see Figure 4 Appendix for analysis using an alternative algorithm that accounts for the influence of the nodes' degree \cite{Kojaku_2018}). The algorithm extends Borgatti and Everett's (\citeyear{borgatti2000}) notion of a single-idealized core to multiple core/periphery structures within a single network. An idealized core/periphery structure is one where core nodes are maximally connected to both other core nodes and periphery nodes, with minimal interconnectivity between the periphery nodes. The algorithm identifies  structures as those that are significantly more similar to the idealized core/periphery model than a random network (S1.4 Appendix). 

Our analyses evaluate the conceptual structure of scientific knowledge in terms of three characteristics—\(C_{it}\), \(R_{it}\), and \(S_{it}\)—which we identified based on prior writings in the philosophy of science. The first parameter, \(C_{it}\), represents the amount of churn or change in the set of core concepts. We measured churn by identifying the fraction of core nodes \(CN\) for each subfield \(i\) that were in the core at year \(t\) and at year \(t-1\) out of the number of core nodes in the core for subfield \(i\)  at year \(t\). We then subtracted the resulting number from 1 to measure the fraction of new concepts introduced to the core for subfield \(i\)  at year \(t\):
\begin{equation}\label{eq:1}
C_{it} = 1 - CN_{it,i(t-1)}/CN_{it}
\end{equation}

The second parameter, \(R_{it}\), measures the relative size of the conceptual cores, defined as the fraction of concepts that comprise the core out of the total number of concepts in the network. We measured the relative size of the cores by calculating the fraction of core nodes \(CN\) for each subfield \(i\) at year \(t\) out of the total number of concepts \(T\)---both core concepts and periphery concepts---for each subfield \(i\) at year \(t\):

\begin{equation}\label{eq:2}
R_{it} = CN_{it}/T_{it}
\end{equation}

The third parameter, \(S_{it}\), represents the number of core/periphery structures for each subfield \(i\) at year \(t\). We calculated the number of cores using Kojaku and Masuda's (\citeyear{kojaku2017}) algorithm, by counting the number of core/periphery structures that are significantly more similar to the idealized core/periphery structure than a random network. 

In supplemental analyses, we explored whether the core/periphery structures we detected could be an artifact of our data by comparing our core/periphery scores to scores calculated using random networks of similar size and configuration (S2.3 Appendix). Our findings show that (1) the core/periphery measures derived from the observed networks differ from those obtained using random networks, (2) the temporal trends we observe in our core/periphery measures persist even after accounting for the expected values of those properties in comparable random networks, and (3) the associations we found between the core/periphery organization of concepts and innovation outcomes are also resilient to adjustment for the expected values of those properties in random networks with similar characteristics.

Finally, using a regression-based approach, we find that our main results hold even after controlling for changes in linguistic and publication practices over time (S2.4 Appendix).

\section{Data availability}
Data from Web of Science and American Physical Society are available from Web of Science and American Physical Society, respectively, but restrictions apply to the availability of these data, which were used under license for the current study, and so are not publicly available. These data are however still available from the authors upon reasonable request and with permission from the respective publishers.

\section{Code availability}
The Python 3, RStudio, and Stata 18 code we used to analyze and visualize the data for the current study will be made publicly available on Zenodo or equivalent platform after acceptance and prior to publication (i.e., after the review team has determined that no additional changes to analyses are required). 

\section{Acknowledgments and Funding Sources}
We thank the National Science Foundation for financial support of work related to this project (grants 1829168 and 1932596). The funders had no role in study design, data collection and analysis, decision to publish or preparation of the manuscript. We also thank Daniel Hirschman, Michael Park, and Yeon Jin Kim for feedback on an earlier version of this work, and Thomas Gebhart for many helpful conversations and assistance with data and computation. Our work was presented as a poster at the 2nd Annual International Conference of Science of Science and Innovation, as a poster at the 43rd Annual Meeting of the Cognitive Science Society, as a lightning talk at Networks 2021: A Joint Sunbelt and NetSci Conference, and as a poster at the 3rd North American Social Networks Conference.

\section{Author contributions}
The study was conceptualized and designed by K.K., E.L., and R.F. Data analysis was conducted by K.K. and R.F. The manuscript was initially drafted by K.K., E.L., and R.F., with subsequent revisions made by K.K. and R.F.

\section{Competing interests}
The authors declare no competing interests.

\pagebreak

\include{innovation_table}

\pagebreak

%TC:ignore

\pagebreak

%TC:ignore
%TC:endignore

%TC:ignore
%TC:endignore

\pagebreak

%TC:ignore 
% \bibliographystyle{apalike}
\bibliographystyle{unsrtnat}
\bibliography{bibliography}

%TC:endignore

\pagebreak

\section*{Supplementary Materials}
\section*{S1. Materials and Methods}
\subsection*{S1.1 Data}

To map the structure and dynamics of scientific concepts, we drew on two large-scale bibliographic databases: the American Physical Society corpus (``APS data'') and the Web of Science (``WoS data''). The APS data consist of detailed records on the more than 630,000 articles appearing in any of the 16 peer-reviewed research journals published by the American Physical Society (e.g., \emph{Physical Review Letters}). In light of our focus on whether and how the organization of scientific knowledge changes over time, the APS data are particularly attractive for their long duration, spanning nearly 125 years (1893-2017). The WoS data consist of detailed records on more than 64 million papers published in 28,968 journals between 1900 and 2017, and have been widely used in prior, related research \citep{hofstra2020, cao2020idea}. While WoS includes bibliographic records for nearly all fields of science and scholarship, we focus our attention on those published in the social sciences (i.e., as classified by WoS). Examining articles from both the physical sciences (via the APS data) and the social sciences (via the WoS data) allows us to compare whether and how the structure and dynamics of concepts differs between distinctive domains of knowledge.

While the Web of Science (WoS) is a widely used in prior work \citep{Li_2018, Birkle_2020}, the data are not without limitations, which should be kept in mind when interpreting our results \citep{Mongeon_2016, Tennant_2020}. A significant bias is its emphasis on English-language journals and uneven coverage across disciplines. Furthermore, WoS tends to overrepresent research from North America and Europe and underrepresent contributions from regions such as Africa, Latin America, or parts of Asia, creating a regional bias that may affect the perceived trajectory and prominence of social sciences, which can vary on a global scale.

Similar limitations are present in the American Physical Society (APS) data set. The APS encompass a set of journals published by a single publisher, which may not capture the full breadth of research in the physical sciences. This limitation could introduce a selection bias, as it reflects the publication preferences and policies of one organization rather than the wider scientific community. Additionally, the APS data set, like WoS, includes only English-language articles, which restricts the cultural and linguistic diversity of the included research. Such constraints could potentially impact our findings by limiting the representativeness of the data in terms of global scientific output and the diversity of scientific discourse.

As we describe in greater detail below, we create our maps separately for subfields of the physical and social sciences, as prior research suggests different domains of knowledge are likely to exhibit distinctive conceptual organization \citep{cole1995}. For the physical sciences, we assign papers (and their associated concepts) to subfields using Physics and Astronomy Classification Scheme (PACS) codes. Papers in the APS data are assigned between 1 and 5 PACS codes, corresponding to their subject matter. PACS codes are hierarchical, with more than 7,300 codes at the most granular level (e.g., 04.30.-w, ``Gravitational waves'') and 10 codes at the highest level of abstraction (e.g., 20, ``Nuclear Physics''). We use these highest level fields to group papers into 10 subfields of the physical sciences (based on the first/primary PACS code listed for each paper). For the social sciences, we assign papers (and their associated concepts) to subfields using subject areas from the Web of Science, which correspond to generally recognized academic disciplines (e.g., ``Sociology''); across all fields of science and scholarship, WoS track around 153 subject areas, 24 of which fall within the social sciences. In contrast to PACS codes, subject areas are assigned to journals (not papers); each journal may be assigned to multiple subject areas (with no designation for a primary subject), and therefore a single paper can contribute concepts to multiple subfields. 

For our analyses, we limit our attention to the period from 1965-2016 and 1992-2016 for the APS and WoS data, respectively. While the APS data span a longer time period, PACS codes were not developed until 1970, and are not assigned reliably in our data until 1980. To impute historical PACS domains, we first vectorized the abstracts of each APS paper through a TF-IDF embedding. We then constructed a training dataset consisting of pairs of paper embeddings and their corresponding PACS domain assignment. This dataset of pairs defines a categorical supervised learning task for which we trained a Random Forest classifier using 3-fold cross-validation and with class weights assigned in inverse proportion to the within-fold class frequency. We used the resultant trained classifier to predict the PACS domains for historical papers which had no assigned PACS domain. We were reluctant, however, to use this approach to classify papers published long before the adoption of PACS codes by the APS, as contemporary subjects in the physical sciences may not map meaningfully onto earlier eras. For the WoS data, our choice to begin in 1992 was based on our decision to extract concepts from paper abstracts; WoS only began reliably recording abstracts in 1992. Finally, we end our analyses in 2016 because that is the last year that we have complete data for both APS and WoS. After subsetting and removing articles without abstracts, our analysis is based on a total of 2,605,224 papers, with 582,679 and 2,022,545 coming from the APS and WoS data, respectively.

\subsection*{S1.2 Concept Extraction Process}
We began the concept extraction process by pulling 636,294 articles from the American Physical Society corpus and 7,200,929 articles from the Web of Science. First, we removed articles that did not contain abstracts, leaving us with 582,679 APS articles and 2,022,545 WoS articles. Concepts were extracted from each abstract using the Textacy package and the spaCy Python package \citep{dewilde2021,honnibal2017}. First, we processed each abstract by removing html tags, white spaces, and line breaks. Next, we used part-of-speech tags to parse the abstracts into single-word and multi-word noun phrases. We then processed each token in the noun phrases by removing instances that only contained spaces, stop words (i.e., a, an, and, another, any, both, enough, he, her, him, his, I, it, its,  many, me, most, much, my, or, other, our, she, some, that, the, their, them, these, they, this, those, us, we, what, whatever, which, whichever, whose, you, your), and punctuation (i.e., '!'\#\$\%\&',.:;?@][`\_{}'~()-). Finally, each token was lemmatized. The processed tokens from multi-word noun phrases were rejoined, and duplicate noun phrases from each abstract were dropped. We then generated a subfield $\times$ year $\times$ article $\times$ noun phrase panel, and dropped rows that were missing noun phrases, had noun phrases without letters, or had noun phrases that were only two or fewer characters long. We also removed punctuation (i.e., '!'\#\$\%\&',.:;?@][`\_{}'~()-”/) and capital letters from the remaining noun phrases. 

\subsection*{S1.3 Edge Weights for Semantic Network Analysis}

Our networks are bipartite, consisting of both concept nodes and article nodes. We conducted our analysis on the concept nodes after projecting the bipartite graph onto the concept layer. We weighted the edges of our network using a common approach, ensuring that the concept networks derived from the bipartite network retained the information of the original bipartite structure. To do this, we weighted the projection of the bipartite graph \( B \) onto the concept node sets using the ``collaboration weighted projection'' method \citep{Newman_2001, Kojaku_2018}. The weight \( W(u, v) \) of an edge between two concept nodes \( u \) and \( v \) in the projected network is calculated as follows:

\[
\text{W}(u, v) = \sum_{k} \frac{\delta_{u}^{k} \delta_{v}^{k}}{d_k - 1},
\]

\noindent where, \( u \) and \( v \) are concept nodes, and \( k \) indexes the article nodes in the bipartite structure. \( d_k \) denotes the degree of the article node \( k \), which is the number of concepts to which the article is connected. The term \( \delta_{u}^{k} \) is 1 if the article node \( k \) is connected to the concept node \( u \) in the bipartite graph, and 0 otherwise. This results in a weighted projection where the concepts are more strongly connected if they are shared by a larger number of articles, with the connection strength adjusted by the number of concepts associated with each article.

\subsection*{S1.4 Detection of Core/Periphery Structures}
To identify core/periphery structures within our concept networks, we utilized the KM-ER algorithm developed by Kojaku and Masuda (\citeyearpar{kojaku2017}). This algorithm was selected based on its suitability for large-scale networks and its ability to identify multiple core/periphery structures within a single network. Our choice of algorithm  was also reinforced by a qualitative assessment: the KM-ER algorithm consistently yielded core/periphery structures that were topically coherent and substantively meaningful, outperforming other methods we considered. This approach extends the widely-used single core/periphery structure detection method of Borgatti and Everett (\citeyearpar{borgatti2000}), (which identifies a single core/periphery structure within a network) by allowing for the detection of multiple core/periphery structures. 

In this section, we provide a brief overview of the algorithm; additional details are given in the original paper \citep{kojaku2017}. Given a graph $G$ with $N$ nodes and $M$ edges, let \textbf{A} = ($A_{ij}$) be an adjacency matrix, wherein $A_{ij} = 1$ when node $i$ and $j$ share an edge and $A_{ij} = 0$ otherwise. Further, let $\mathbf{x} = (x_{1},x_{2},...,x_{N})$ represent a vector of length $N$, where node $i$ is a periphery node if $x_{i} = 0$ and a core node if $x_{i} = 1$. An idealized core/periphery structure is defined as a network wherein each core node is adjacent to all other core and periphery nodes, and each periphery node is adjacent to all core nodes and not adjacent to any periphery nodes. To extend this notion to multiple core/periphery structures, let $C$ be the number of core/periphery structures and $\mathbf{c} = (c_{1},c_{2},...,c_{N})$ be a vector of length $N$, where $c_{i}  \in \{1,2,\ldots , C\}$ indexes core/periphery structures that include node $i$. Overlap is prohibited, such that  each node $i$ is associated with only one core/periphery structure, and is either assigned to the core or periphery (but never simultaneously both) within that structure. The corresponding adjacency matrix, $\mathbf{B} (\mathbf{c,x})$, is defined as follows 

\begin{equation}
    B_{ij}(\mathbf{c,x}) = \begin{cases}
    \delta_{c_{i},c_{j}}, & (x_{i} = 1 \text{ or } x_{j} = 1,\text{ and } i \neq j),\\
    0 & \text{(otherwise)},
    \end{cases}
    \end{equation}
    
where $\delta$ is Kronecker delta. The algorithm searches for the $(\mathbf{c,x})$ that results in $\mathbf{B} (\mathbf{c,x})$ that is the most similar to $\mathbf{A}$ by maximizing 

\begin{equation}
    \begin{gathered}
    Q^{cp}(c,x) = \sum_{i=1}^{N}\sum_{j=1}^{i-1}A_{ij}B_{ij}(c,x) - \sum_{i=1}^{N}\sum_{j=1}^{i-1}pB_{ij}(c,x) \\
     = \sum_{i=1}^{N}\sum_{j=1}^{i-1}(A_{ij}-p)(x_{i}+x_{j}-x_{i}x_{j})\delta_{c_{i},c_{j}}.
    \end{gathered}
    \end{equation}    
    
The density of edges in the network is denoted by $p$, where $p = M/[N(N-1)/2]$. $\sum_{i=1}^{N}\sum_{j=1}^{i-1}A_{ij}B_{ij}(c,x)$ represents the number of edges found in both the observed network and the idealized core/periphery structure. $\sum_{i=1}^{N}\sum_{j=1}^{i-1}pB_{ij}(c,x)$ captures the number of edges found in both the idealized core/periphery structure and an Erd\H{o}s-R\'{e}nyi random graph, where the adjacency between two nodes is based on probability $p$. $Q^{cp}$ indicates whether the observed network and the idealized core/periphery structure have more edges in common than chance, and ranges between $-M$ and $M$. Kojaku and Masuda (\citeyearpar{kojaku2017}) introduce a label switching heuristic to identify $Q^{cp}$ with the largest value; we use this heuristic to maximize $Q^{cp}$ in our study. We ran the algorithm 10 times and computed the average across both networks and runs to obtain values for each core/periphery measure.

\subsection*{S1.5 Creation of the Consensus Dictionary}
To measure scientific consensus, we derived a dictionary of words that expressed consensus (e.g., agree, compatible, support) through a sequence of stages. We began by having three independent raters each review 50-70 abstracts from our data that included in-text citations, and identify the words used to express consensus with cited findings. For example, the following statement from an abstract in our data uses ``agreement'' to indicate consensus: ``We find the number of fingers observed in our simulations to be in excellent agreement with experimental observations and a linear stability analysis reported recently by Smolka and SeGall (2011)'' \citep{mayo_2013}. We only analyzed physical sciences abstracts from the APS data because the database uniformly denoted in-text citations using XML tags. The social sciences abstracts from the WoS data did not have a unified indicator of in-text citations, preventing us from extracting the appropriate abstracts for further analyses. The raters then compared the consensus-related words identified and resolved discrepancies. Next, we used our set of abstracts to train word embeddings, and pulled out the 5 most similar words to each word identified by the raters. The three raters were then given the complete list of consensus words---containing both words identified by the raters and derived through word embeddings---and asked to perform a forced-choice comparison between pairs of words, choosing the term that was most indicative of consensus. The final dictionary, reproduced below, only included words above the median ranked score. 

\subsection*{S1.6 Consensus Dictionary}\label{sec:consensusDictionary}
The list below reports the words that were used as indicators of scientific consensus in paper abstracts.
\begin{multicols}{2}
\begin{enumerate}\itemsep0em
\item acceptable
\item accessible
\item accordance
\item account
\item advantageous
\item agree
\item agreement
\item apply
\item archetypal
\item attributable
\item base
\item benefit
\item compatible
\item complement
\item confirm
\item connection
\item consensus
\item consistent
\item consistently
\item correct
\item correctly
\item develop
\item exemplify
\item explainable
\item fit
\item highlight
\item ideal
\item illustrate
\item implement
\item inspire
\item noteworthy
\item perfect
\item reinforce
\item relevant
\item similar
\item substantiate
\item suitable
\item support
\item uniformly
\item unify
\item valid
\item verify
\end{enumerate}
\end{multicols}

\section*{S2. Results}

\subsection*{S2.1 Properties of Scientific Concepts}
In this section, we provide a more detailed discussion of the properties of scientific concepts. We begin our analysis by conducting a series of investigations comparing the properties of scientific concepts that appear in the core versus the periphery (see Table~\ref{table:MechanismValidation}). 

Consistent with intuition, we found that core concepts are generally more established within a scientific field than periphery concepts. Specifically, we measured the age of each concept, the number of subfields, and papers in which they occur, and, for the physical sciences, whether the concept was among the 892 curated topical tags on the Physics Stack Exchange Q\&A site. The results generally suggest that core concepts are relatively more established than periphery concepts. We found that core concepts from the social sciences appeared in more subfields $(\text{Welch's t} = 28.12; p<0.001)$ and were used in more papers $(\text{Welch's t} = 6.11; p<0.001)$. For the physical sciences, core concepts were older than periphery concepts $(\text{Welch's t} = 5.56; p<0.001)$ and were used in more papers $(\text{Welch's t} = 18.42; p<0.001)$. We also found, for physical sciences concepts, that  on average, core concepts are more likely to appear as tags on Stack Exchange than periphery concepts: an average of 7\% for core concepts and 5\% for periphery concepts $(\text{Welch's t} = 12.54; p<0.001)$. See Table~\ref{table:MechanismValidation} for more information. 

Our findings also suggest that periphery concepts have a higher level of specificity when compared to core concepts. We compared core and periphery concepts in terms of the number of words, characters, and digits present in each concept. We expected that periphery concepts would have a higher level of specificity. As theorized by Lakatos (\citeyearpar{lakatos1978}), periphery concepts or auxiliary hypotheses are of higher specificity because they are used to empirically test the abstract/theoretical core. For the social sciences, we found that periphery concepts on average contain a higher number of words than the concepts $(\text{Welch's t} = -15.43; p<0.001)$ and that core concepts had fewer characters than periphery concepts $(\text{Welch's t} = -19.70; p<0.001)$. For the physical sciences, we found that periphery concepts on average contain a higher number of digits than core concepts $(\text{Welch's t} = -8.78; p<0.001)$ and that core concepts had fewer characters than periphery concepts $(\text{Welch's t} = -9.85; p<0.001)$. In general, our findings suggest that periphery concepts involve a higher level of specificity when compared to core concepts, as expressed by an increase in the number of words, characters, and digits. See Table~\ref{table:MechanismValidation} for more information. 

We also found that core concepts at year $t$ were more likely to remain in the core at year $t+1$, whereas the likelihood that periphery concepts at year $t$ remained in the periphery at year $t+1$ increased over time (see Figure~\ref{fig:mobility_concepts}). First, we looked at how core concepts changed over time. We calculated the proportion of core concepts at year $t$ that remained in the core, transitioned to the periphery, or exited the network at year $t+1$. We found that the majority of core concepts remained in the core for both the social sciences (between 54\% and 63\%) and the physical sciences (between 42\% and 61\%). We also found that core concepts were more likely to transition to the periphery than exit the network for the majority of years in both the social and the physical sciences. Next, we looked at how periphery concepts changed over time. We calculated the proportion of periphery concepts at year $t$ that remained in the periphery, transitioned to the core, or exited the network at year $t+1$. We found that the likelihood that periphery concepts remained in the periphery increased over time. Additionally, if periphery concepts exited the periphery, they were more likely to transition to the core than exit the network for all years in both the social sciences and the physical sciences. 

\subsection*{S2.2 Robustness Checks and Validation of Findings}

To ensure the robustness of our findings, we conducted a series of supplementary analyses. Specifically, we evaluated the consistency of our results using a range of different thresholds, an alternate algorithm, a three-year moving time window, and among distinct subfields. Across all these methodologies, we consistently observed congruent results, reaffirming the reliability of our primary findings.

\subsubsection*{S2.2.1 Alternative Thresholds and Algorithm}

To assess the consistency of our findings across various thresholds, we conducted further analyses by retaining concepts that appeared in more than a specified minimum percentage and less than a designated maximum percentage of abstracts. Specifically, we conducted our analysis across 25 distinct thresholds, encompassing every combination of our minimum percentage values (0.00\%, 0.001\%, 0.025\%, 0.05\%, and 0.1\%) and our maximum percentage values (1\%, 5\%, 7.5\%, 10\% and 100\%). For the analysis reported in the main manuscript, we focused on the findings related to the threshold yielding the highest average $Q^{cp}$-value, which retained concepts that occurred in a minimum of 0.1\% of abstracts and a maximum of 1\% of abstracts. The $Q^{cp}$-value serves as a quality metric, quantifying the extent to which a particular network partition conforms to the ideal core/periphery network. Furthermore, upon substantive examination of the core/periphery structures identified, the selected threshold yielded results that were deemed most reasonable for the KM-ER algorithm.

As depicted in Table \ref{fig:over_time_multi_thresholds}, most threshold combinations exhibit a consistent trend over time in both the social and physical sciences. The trends indicate a decline in both the churn in core concepts and the relative number of core nodes. We find somewhat mixed results for the progression of the number of cores over time for the social and physical sciences.  In the social sciences, the data predominantly displays stagnant trends, albeit with occasional minor increases and decreases. Only a few thresholds exhibit a discernible upward trajectory. This variability indicates that the number of cores may be contingent upon the chosen threshold, making it difficult to reach a definitive conclusion. In the main manuscript, we observe a marginal decrease in the number of cores. However, the newly reported results imply that such a trend is not consistently observed across multiple thresholds. In contrast, the physical sciences demonstrate a more coherent pattern, with most thresholds suggesting an overall increase in the number of cores over time. Additionally, we observed that concepts, regardless of the threshold, become increasingly centralized within a limited number of core/periphery structures as time progresses. This consistency implies that our conclusions remain supported across different thresholds. 

To ensure the robustness of our findings, we cross-validated our analysis using an alternate core/periphery detection algorithm. This algorithm, referred to as KM-config \citep{Kojaku_2018}, identifies multiple core/periphery structures while better accounting for heterogeneity degree distribution of the nodes in the networks. More specifically, one concern with the KM-ER algorithm used in the main text is that the core/periphery structure of the underlying networks may be driven by node degrees, such that high-degree nodes are consistently labeled as core, while low-degree nodes as peripheral. KM-config introduces a scalable algorithm for identifying pairs of core and periphery, accounting for the influence of node degrees by using a configuration model rather than an Erd\"{o}s-Renyi random graph as the null model. Using this alternative method, we are able to further validate our findings, by showing that they hold when the core/periphery detection algorithm accounts for heterogeneous degree distributions.

As illustrated in the Figure \ref{fig:over_time_km_config}, the outcomes using KM-config generally parallels those presented using KM-ER, highlighted in the main manuscript. For both the social sciences and the physical sciences, the plots show a decrease in the churn of core concepts and a decrease in the relative number of core nodes (across most thresholds). For the social sciences, the results for the number of cores paints a complex picture: while numerous thresholds show a stable trend with a minor decrease, mirroring the main manuscript's findings, there are also instances where an increase in the number of cores is evident. By contrast, the physical sciences exhibit a trend that aligns more consistently with the main manuscript, with the majority of thresholds demonstrating an overall upward trajectory in the number of cores over time. The inset plots show that concepts become increasingly concentrated in a few number of core/periphery structures across time. However, we also find that KM-config is more sensitive to the choice of threshold, specifically with respect to the relative size of the core. We see that a portion of the results show that the relative size increases over time for the social sciences (10 out of 25 thresholds) and for the physical sciences (2 out of 25 thresholds). However, the majority of the thresholds across the sciences support the trend that the relative number of core nodes are decreasing over time.

\subsubsection*{S2.2.2 Three-Year Moving Window}

We extend our initial analysis---which examines year-to-year network variations---by assessing how the core/periphery parameter evolves within a three-year rolling window (see Figure \ref{fig:3year_moving_window}). Our findings largely align with those presented in the main manuscript's year-to-year analysis. Specifically, the results indicate a consistent decline over time in the churn of core concepts and the relative size of the cores in both the social and physical sciences. In the social sciences, the number of cores remains relatively stable with a slight decline, which is in line with the results reported in the main manuscript. In the physical sciences, we observe an initial increase in the number of cores, which then levels off in recent years, echoing our main observations. However, when examining the results through a three-year moving window, the peak in the physical sciences appears earlier. In line with our primary findings, we also see a growing concentration of core concepts within a single core/periphery structure in both fields. These outcomes affirm the stability and reliability of our research conclusions.

\subsubsection*{S2.2.3 Distinct Subfields}

Lastly, we examined the temporal trends of individual subfields. While we primarily focus our analysis on the overarching trends across various fields, examining each subfield independently provides insight into any significant deviations from the aggregate results we have reported. In general, the temporal trends observed within individual subfields align with the collective findings (See Fig \ref{fig:over_time_subfields}). Both for the social sciences and physical sciences, our results depict a decline in the churn of core concepts and in the relative number of core nodes. We also see a slight decline in the number of cores for the social sciences. For the physical sciences, certain subfields correspond more closely to the results reported in the main manuscript; in general, we see an initial surge in core numbers, which peaks before the end of the observed period. The inset plots indicate that concepts increase in concentration around a limited set of core/periphery structures over time.

\subsection*{S2.3 Comparison of Observed Networks with Similar Random Networks}

To investigate whether the core/periphery structures identified could be artifacts of the underlying data from which they are constructed, we benchmarked them against comparable random networks. For each observed network, we generated 50 random networks, preserving the degree distribution of the original bipartite networks. This randomization was designed to keep the number of papers linked to each concept constant, as well as the number of concepts mentioned in each paper, ensuring that the structural integrity of the network was maintained.

\paragraph{Generation of Null Models}
Our null models began with the original bipartite network, consisting of paper nodes connected to concept nodes. We employed a shuffling algorithm, commonly referred to in the literature as a degree-preserving edge swap, to randomize the connections while preserving the degree of each paper node \citep{uzzi2013atypical,Cimini_2022}. This approach not only ensures that the overall structure of the network remains intact but also controls for variations in abstract lengths and the exponential growth in the number of papers over time, which could otherwise affect the density of the projected concept network.

\paragraph{Projection and Measurement}
Following randomization, we projected these bipartite networks onto a single layer of concept nodes, collapsing multiple edges into single connections. As with the observed networks, we weighted the edges in the randomized networks in proportion to the number of papers on which the adjacent concepts appear together \citep{Newman_2001,kojaku2019multiscale}. For these randomized concept networks, we computed the same core/periphery measures applied to the observed networks: churn of core nodes, relative size of the cores, and the number of cores.

\paragraph{Comparative Analysis}
To assess the significance of the core/periphery properties in the observed networks, we compared them against the distributions generated from the random networks. We calculated a comparative index for each observed network property using the following formula:

\[
\text{Comparative Index} = \frac{\text{Observed Value} - \text{Mean of Random Network Values}}{\text{Standard Deviation of Random Network Values}}.
\]

This index, mathematically analogous to a z-score, quantifies how much the observed network deviates from what might be expected in a random configuration, given the same year and subfield. A detailed presentation of this analysis is provided in Figure~\ref{fig:random_networks}.

This figure showcases cumulative distributions that highlight the variance in core/periphery structures between observed networks and their randomized equivalents. Consistently, the churn of core nodes in the observed networks was lower than that in their randomized counterparts across both the social and physical sciences. The observed networks were also characterized by relatively larger numbers of core nodes when compared to the random networks. In the physical sciences, the observed networks generally had fewer cores than the randomized ones, while in the social sciences, the number of cores were comparable between the observed and randomized networks. Interestingly, the physical sciences demonstrated a higher degree of concentration in the actual networks versus random ones, suggesting a more cumulative knowledge structure. The social sciences showed a contrasting pattern, which may reflect inherent differences in how knowledge is accumulated across disciplines.

We also conducted an analysis where the core/periphery properties derived from the comparative indexes served as predictors in our inferential models—specifically examining their predictive power regarding innovation, as measured by Funk and Owen-Smith's (\citeyearpar{funk2017}) CD index. This robustness check yielded findings that are consistent with the results from our original models in Table \ref{table:RandomNetworkRegressions}. Specifically, we find that disruptive innovation is positively and significantly predicted by the churn of core concepts (\(\textrm{social:} \beta = 0.0034, SE = 0.0002, p < 0.001; \textrm{physical:} \beta = 0.0025, SE = 0.0004, p < 0.001\)) and the relative size of the cores (\(\textrm{social:} \beta = 0.0014, SE = 0.0002, p < 0.001; \textrm{physical:} \beta = 0.0022, SE = 0.0003, p < 0.001\)). For the social sciences, innovation was positively and significantly predicted by the number of cores, which matches what was reported in the main manuscript (\(\beta = 0.0058, SE = 0.0002, p < 0.001\)). For the physical sciences, the number of cores is not a statistically significant predictor of disruptive innovation, which is also consistent with our findings in the main manuscript (\(\beta = -0.0010, SE = 0.0006, p < 0.01\)). Overall, these results are supportive of our main findings. Such consistency suggests that the core/periphery properties of the observed networks possess intrinsic predictive value that cannot be solely attributed to the underlying data structure or the potential confounders therein. In other words, the core/periphery properties in the observed networks provide additional information that is relevant for predicting innovation outcomes, beyond what might be ``baked in'' from structural characteristics of the data from which the networks are derived (e.g., the number of concepts per paper).

In addition to our analyses using the comparative indexes, we conducted an analysis using the core/periphery properties of the random networks as control variables. This approach parallels that of our models with the comparative indexes: by adjusting for the core/periphery properties of the random networks, the coefficient estimates for the observed network properties will reflect the association between the core/periphery properties and disruptiveness beyond what is attributable to the structural characteristics of the data from which the networks are derived. Because each observed network is paired with multiple random networks, the number of observations in this analysis is larger than in our main models (standard errors are adjusted accordingly). 

Table~\ref{table:RandomNetworkRegressionsAdjustedB2} shows the results, which are similar to and substantively supportive of our main findings. For both the social and physical sciences, disruptive innovation is positively and significantly predicted by the churn of core concepts (\(\textrm{social:} \beta = 0.2055, SE = 0.0800, p < 0.05; \textrm{physical:} \beta = 0.1953, SE = 0.0424, p < 0.001\)) and the relative size of the cores (\(\textrm{social:} \beta = 0.0047, SE = 0.0005, p < 0.001; \textrm{physical:} \beta = 0.0009, SE = 0.0003, p < 0.01\)). Consistent with our main results, we do not observe a statistically significant relationship between the number of cores and disruptive innovation in the physical sciences. The only result that deviates from our main models is the relationship between the number of cores and disruptive innovation in the social sciences, where we now see a negative and statistically significant association (\(\beta = -0.0002, SE = 0.0001, p < 0.05\)). However, we note that the coefficient estimate is comparatively small and, more importantly, is consistent with our observation elsewhere in the manuscript that relative to the other core/periphery properties, the number of cores has a less clear temporal trend and relationship with innovation outcomes. 

Finally, we note that the coefficient estimates for the random network core/periphery properties are typically much smaller (often an order of magnitude) than those of the observed networks, and except for the number of core nodes for the social sciences, none are statistically significant. In short, this analysis adds further confidence that our results are not driven by data artifacts (e.g., number of concepts per paper).

\paragraph{Analysis of Over Time Trends}

We also investigated whether the trends we observe in core/periphery organization could be influenced by structural properties that are present in both the observed and random networks. For instance, these trends could be related to dynamic changes in linguistic properties, such as changes in language use and abstract length, or publication rates. To evaluate this possibility, Figure~\ref{fig:R_adjusted_B2} reports the results of analyses that examines changes in core/periphery properties over time after adjusting for expected properties in the random networks. We make these adjustments using a regression-based approach. Specifically, we estimated regression models wherein the dependent variable was the core/periphery property of interest in the observed network and the values of the core/periphery property in random networks were included as a control. Our models also included dummy variables for each year in our data and fixed effects for subfields of the social and physical sciences. The plots in Figure~\ref{fig:R_adjusted_B2} depict the conditional yearly mean of a specific core/periphery property obtained from these regression models. 

As shown in Figure~\ref{fig:R_adjusted_B2}, our results are generally similar after accounting for the baseline effects expected from the structural characteristics of randomly shuffled networks, suggesting that the changes we observe in the conceptual structure of scientific knowledge over time are not fully explained by mere alterations in the distribution of concepts or the number of papers per concept. For both the social and physical sciences, there is a downward trend in the relative number of core nodes ($R_{it}$). The trends in the number of cores ($S_{it}$) also align with those found in our primary analysis. In the social sciences, there is a modest decline in $S_{it}$ over time; for the physical sciences, the number of cores initially increases, but then levels off and begins to decrease. Trends in core concentration, as shown in the inset plots of Figure~\ref{fig:R_adjusted_B2}, panels c and f, further echo our main findings. With respect to the churn of core nodes ($C_{it}$), we consistently find evidence of a downward trend in both fields, in line with our primary results. Notably, in the social sciences, there is a brief reversal at approximately 2005, followed by a leveling off around 2010.

While the trends in core churn mirror our main findings, the magnitude of the decline is less pronounced after adjusting for what might be expected in a random network configuration. Although this result could be interpreted as suggesting that the decline we observe in the churn of core nodes in our main analyses is an artifact, we believe that the attenuation in the downward trend is most likely the result of a combination of how we quantify core churn and the nature of the null model. Our null model adjusts the connections between papers and concepts and, ultimately, after projection, the connections between concepts themselves. However, the identity of the concepts is preserved by the random shuffling procedure. Concepts that appear in more abstracts and consistently across adjacent years are disproportionately likely to be among the core nodes, even after redistributing those concepts across abstracts. Consequently, this particular null model may not be ideal for assessing churn, as it only modifies relational ties and not the identity of nodes. The core/periphery structures of the observed and randomized networks could differ significantly in terms of number, size, and other properties (and based on the distributions of the other core/periphery properties, they do), but in both cases, concepts that appear across more abstracts are likely to be among those labeled as core. Moreover, within a field, popular concepts are also disproportionately likely to appear across adjacent years. Therefore, given the way churn is calculated (based on the overlap in core nodes between adjacent time periods), the null model's shuffling of concept-abstract pairs within field and year is likely to result in a fair degree of consistency in the persistent core nodes.

We evaluated this possibility using an alternative null model. Like our main null model, we randomly shuffled/redistributed concepts across papers, maintaining the same number of papers per field/year and the same number of concepts per paper. However, in this alternative model, we allowed the shuffling to occur across fields (but within years). This approach is likely to break the correlation between adjacent years in terms of the identities of the most connected concepts. By permitting concepts to shuffle across papers not just within a single field but across all fields within the same year, this model significantly increases the 'landing spots' for any given concept. With this broader pool of concepts to draw from, the probability that the same concepts will appear as core in subsequent years by random chance is reduced. Consequently, this adjustment should result in a more noticeable difference in core churn when compared to the main null model, which only shuffles within fields and may inadvertently maintain some year-to-year consistency in core concept identities due to a smaller pool of concepts. Moreover, because the shuffling is performed within year but across fields, it should account for general changes in linguistic patterns over time that affect all fields equally. We anticipate that in this alternative model, we will observe trends in the relative size of the core, number of core nodes, and core concentration that are similar to our main null model. However, we expect a change in the trend of churn over time after accounting for the null model, with the magnitude of the trend showing a more pronounced decline. The results, shown in Figure~\ref{fig:R_adjusted_FB2}, are consistent with this prediction; here, we see a persistent decline in core churn that is comparable in magnitude to our main results. Note that across the two null models, the trends in the other core/periphery properties remain similar after adjustment (i.e., the main difference is in the result for core churn), which adds further support to our interpretation above.

\subsection*{S2.4 Adjustment for Changes in Linguistic and Publication Practices}

In addition to our simulation-based approach, we also evaluated the potential effect of changes in linguistic and publication practices on our results using regression. Specifically, we estimated a series of regressions predicting the core/periphery properties---churn of core nodes, relative number of core nodes, number of cores, and core concentration---for each field $\times$ year in our sample after controlling for field $\times$ year level measures of the average \emph{Number of authors per paper}, \emph{Number of citations made per paper}, \emph{Length of sentences in paper abstracts}, \emph{Number of sentences in paper abstracts}, \emph{Length of words in paper abstracts}, \emph{Number of words in paper abstracts}, \emph{Fraction of nouns in paper abstracts}, \emph{Fraction of adjectives in paper abstracts}, \emph{Fraction of verbs in paper abstracts}, and \emph{Fraction of adverbs in paper abstracts}. In addition, our models included fixed effects for each subfield and year. Results of these analyses are shown in Figure~\ref{fig:language_controls_adjusted}; values correspond to the conditional mean of each respective core/periphery property for each year (estimated using year dummy variables). 

Overall, these analyses provide strong support for our main findings. Even after adjusting for changes in linguistic and publication practices, we continue to find comparable or steeper downward trends in the churn of core nodes and relative number of core nodes and upward trends in core concentration across both the social and physical sciences. For the physical sciences, trends in the number of cores are also similar to our main results after adjustment. For the social sciences, adjustment for linguistic and publication practices results in a modest increase in the predicted number of cores over time. However, the increase in the number of cores shown in our main results and the decrease shown here are both modest with respect to magnitude, which may, when taken together, suggest that the number of cores in the social sciences is comparatively stable.

\pagebreak

%TC:ignore 
% \bibliographystyle{apalike}
% \bibliographystyle{unsrtnat}
% \bibliography{bibliography}

%TC:endignore

\pagebreak

\begin{figure}[ht!]
\centering
\includegraphics[width=0.75\textwidth]{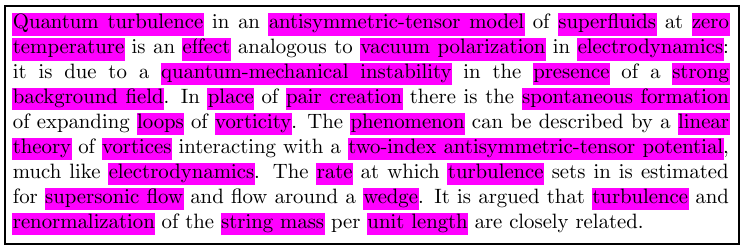}
\caption{\textbf{Concepts extracted from the text of an abstract.} This figure shows an example abstract from the APS data; the highlighted text indicates single-word and multi-word noun phrases identified as concepts using our extraction algorithm. }
\label{fig:parsing}
\end{figure}

\pagebreak
\begin{figure}[ht!]
\centering
\includegraphics[width=\textwidth]{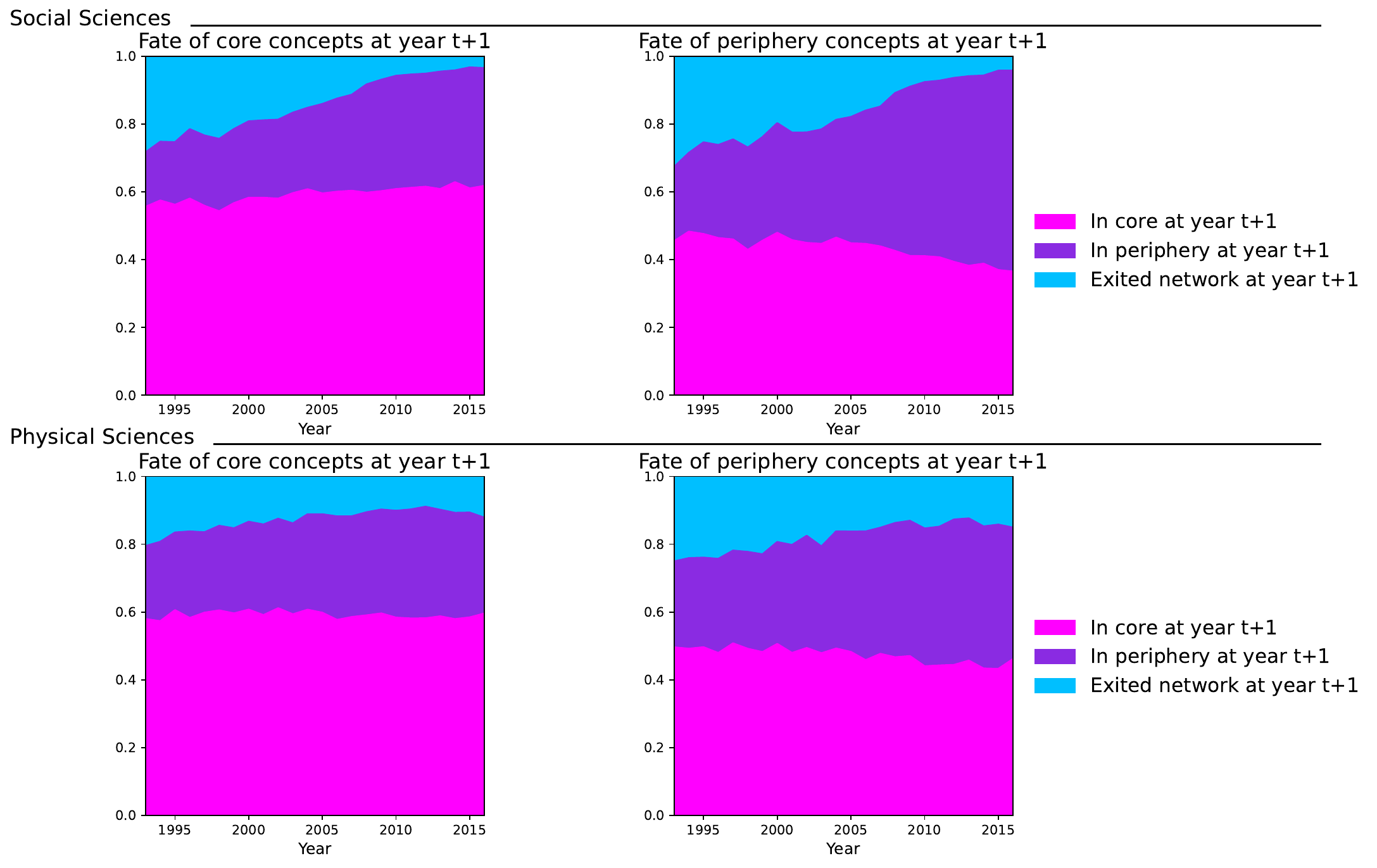}
\caption{\textbf{Mobility of concepts to/from the core/periphery over time.} To generate this figure, we calculated the proportion of core nodes at year $t$ that either stayed in the core, transitioned to the periphery, or exited the network at year $t+1$. Our results show that core concepts are more likely to stay in the core at year $t+1$ for both the social sciences (top, left) and physical sciences (bottom, left). Further, we find that if core concepts exit the core at year $t+1$, they are more likely to enter the periphery than exit the network. Next, we calculated the proportion of periphery nodes at year $t$ that either stayed in the periphery, transitioned to the core, or exited the network at year $t+1$. Our results show that periphery concepts at year $t+1$ are more likely to stay in the periphery as time progresses for the social sciences (top, right) and for the physical sciences (bottom, right). We also observe that if periphery concepts exit the periphery, they are more likely to enter the core than exit the network.}
\label{fig:mobility_concepts}
\end{figure}

\pagebreak

\begin{figure}[ht!]
\centering
\includegraphics[width=1\textwidth]{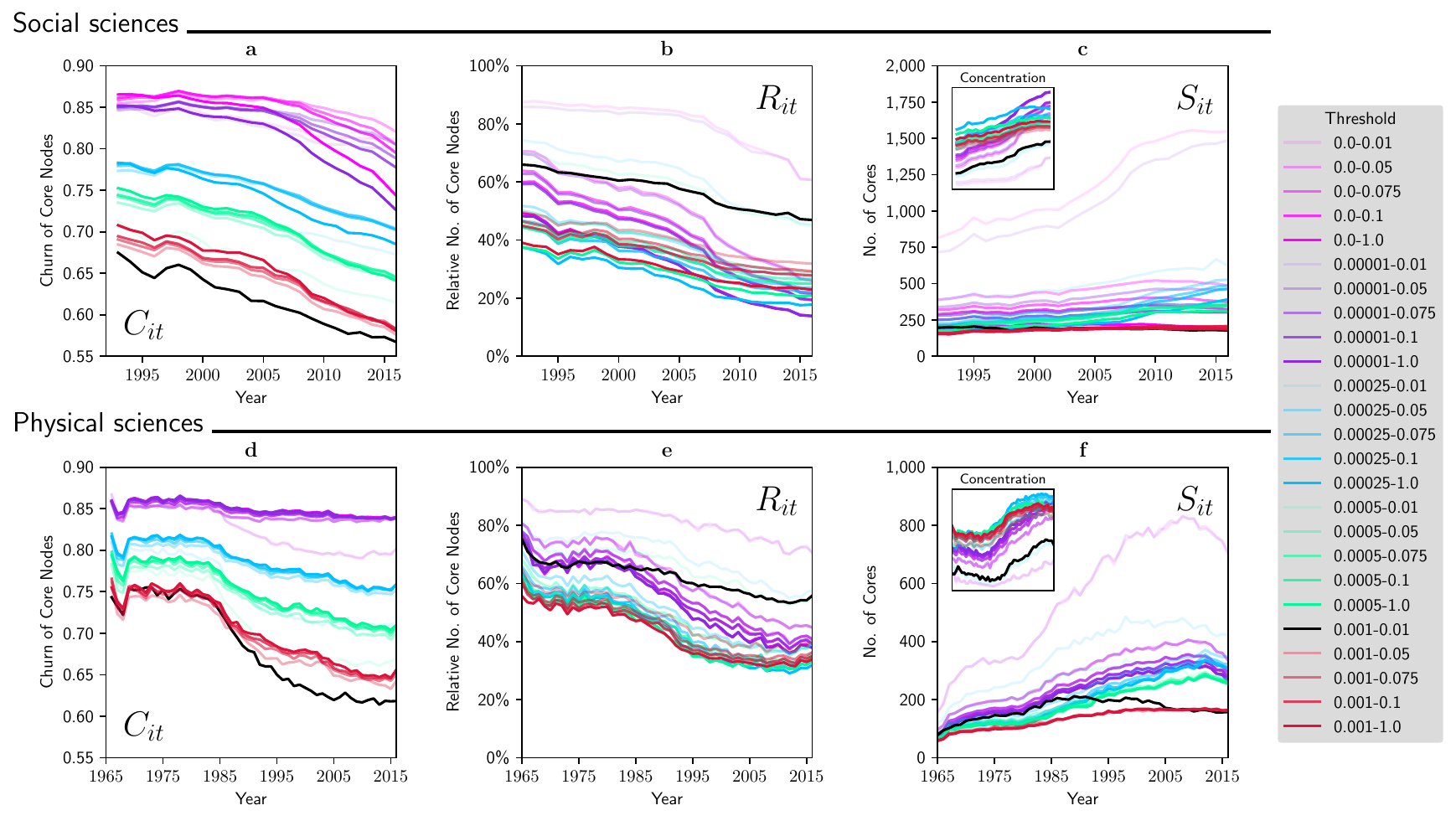}
\caption{\textbf{Conceptual structures over time using multiple thresholds.} The plot shows changes in the core/periphery organization averaged across all of the subfields for 25 different thresholds. The over time trends are similar to the ones reported in the main manuscript, where we kept concepts that occurred in more than 0.1 percent and fewer than 1.0 percent of abstracts. Specifically, we see a general decline for both the social sciences and the physical sciences over time in terms of the churn in core concepts (a,d) and the relative size of the cores (b,e). For the social sciences, we see mixed results for the number of cores, where some thresholds seem relatively flat while others clearly increase (c). However, the results for the physical sciences consistently show an overall increase in the number of the cores over time (f). Additionally, we find an increase in the concentration of concepts within a single core/periphery structure, which is calculated using the Herfindahl index (see inset plots in c and f). An increase in the concentration of concepts suggest that the largest core is getting bigger, which is consistent with the findings reported in the main manuscript. }
\label{fig:over_time_multi_thresholds}
\end{figure}
\pagebreak

\begin{figure}[ht!]
\centering
\includegraphics[width=1\textwidth]{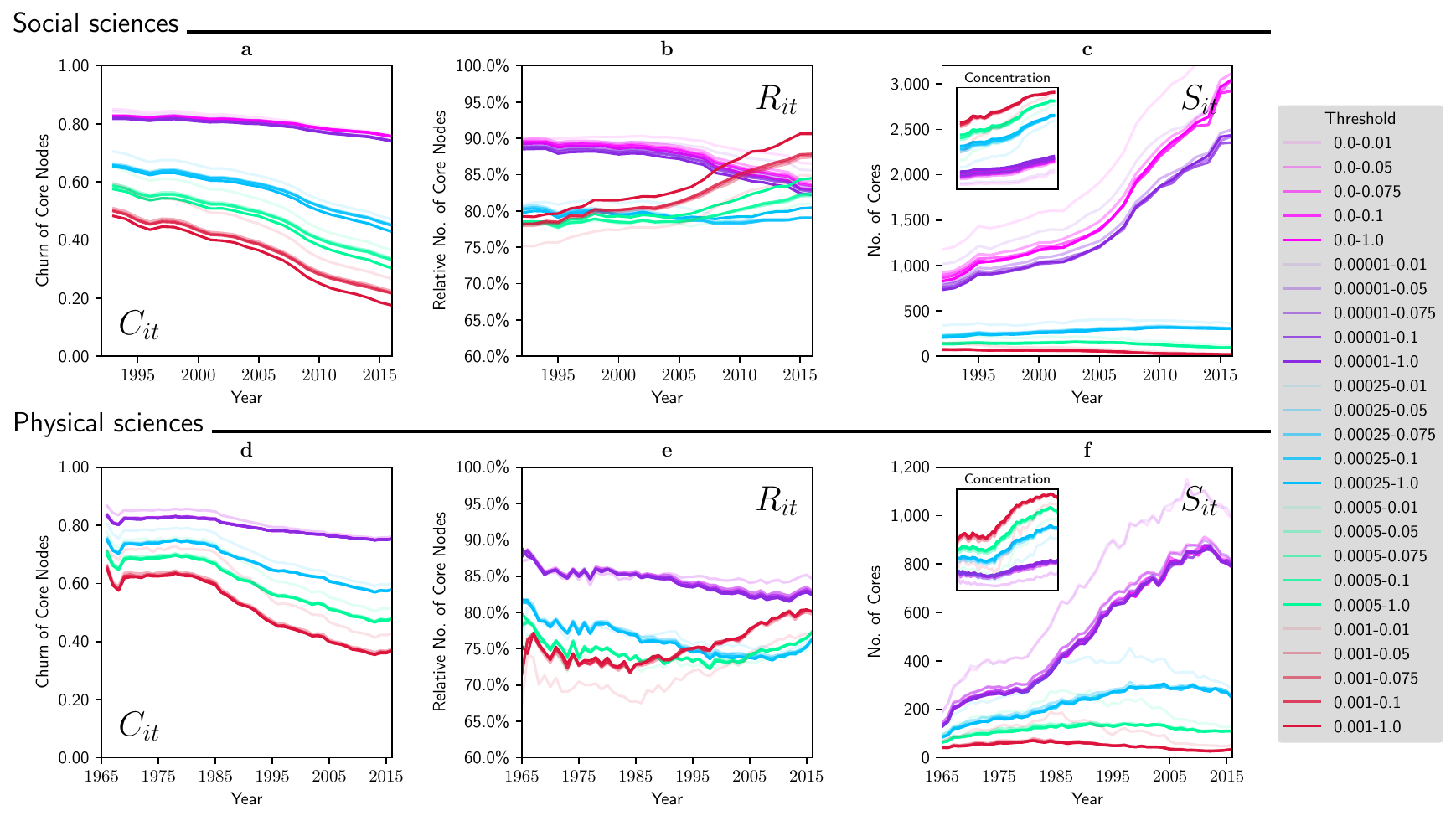}
\caption{\textbf{Conceptual structures over time measured using an alternative core/periphery detection algorithm.} The plot shows changes in the core/periphery organization averaged across subfields measured using the KM-config core/periphery algorithm \citep{Kojaku_2018}. KM-config is similar to the KM-ER algorithm used for our main analysis, but uses a configuration model as a null model rather than an Erdos-Renyi random graph. For most thresholds, results using the KM-config algorithm are similar to those reported in the main text. For both the social sciences (top row) and physical sciences (bottom row), the plots show a decrease in the churn of core concepts (a,d) and a decrease in the relative number of core nodes (b,e) (across most thresholds). For the number of cores, we find mixed results for the social sciences (c) and an overall increase in the number of cores (again across most thresholds) for the physical sciences (f). The inset plots (found in plots c and f) show that concepts become increasingly concentrated in a few number of core/periphery structures across time. However, we also find that KM-config is more sensitive to the choice of threshold, specifically with respect to the relative size of the cores.}
\label{fig:over_time_km_config}
\end{figure}

\pagebreak

\begin{figure}[ht!]
\centering
\includegraphics[width=\textwidth]{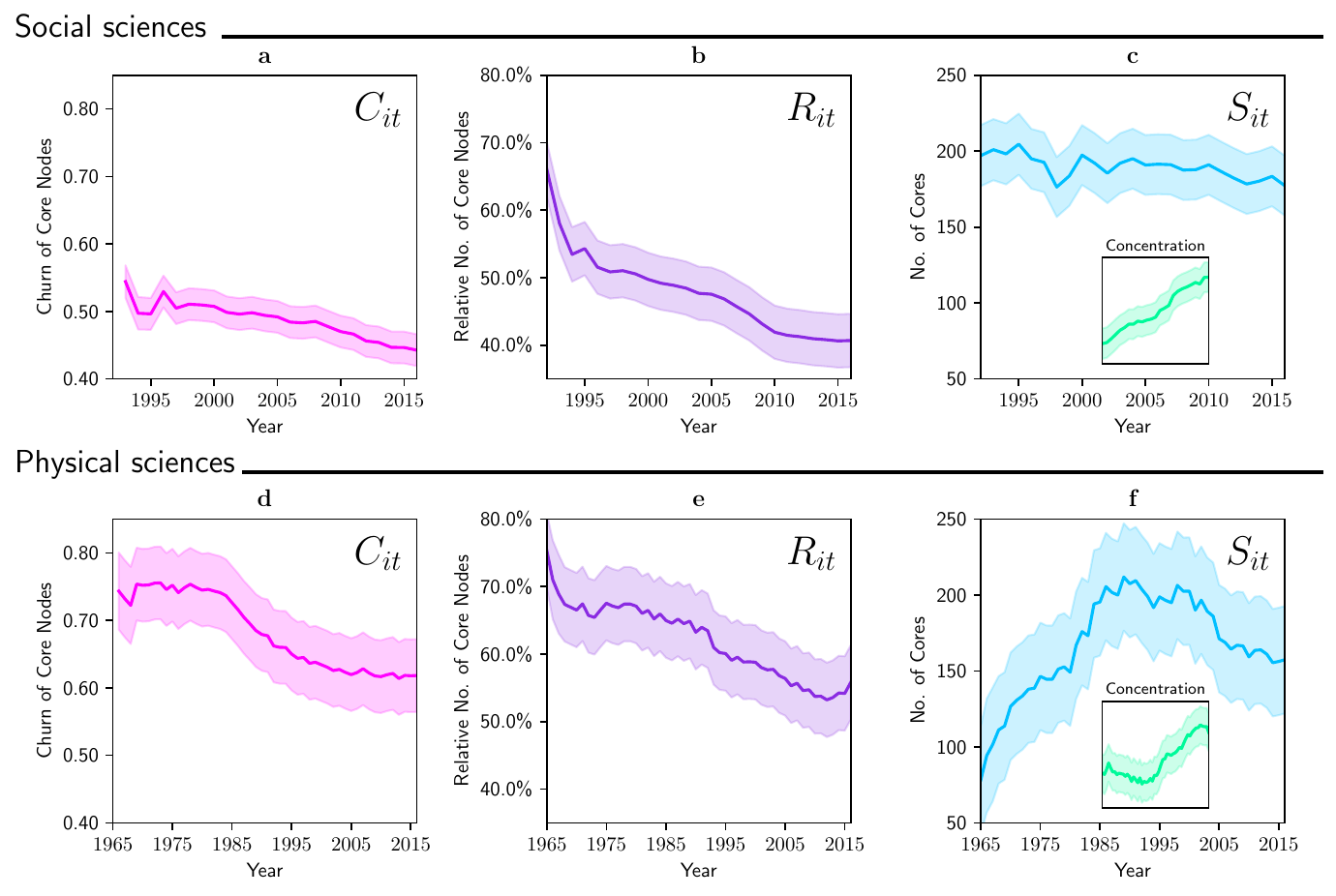}
\caption{\textbf{Over time plots of core/periphery measures using a 3-year moving window.} We find similar results to the ones reported for the year-to-year analysis in the main manuscript. Specifically, these results suggest that the churn of core concepts and the relative size of the cores decrease over time for both the social sciences and the physical sciences. For the social sciences, the number of cores is more or less stable over time, consistent with our main results; for the physical sciences, we observe an increasing number of cores followed by a flattening-out phase in recent years, similar to our main results, although for the plots using the 3-year moving window, the peak comes earlier. Consistent with our main results, the concentration of core concepts around a single core/periphery structure also increases for both the social sciences and physical sciences (see inset plots on top right and bottom right figures). Overall, these results indicate the robustness of our findings.}
\label{fig:3year_moving_window}
\end{figure}

\pagebreak

\begin{figure}[ht!]
\centering
\includegraphics[width=1\textwidth]{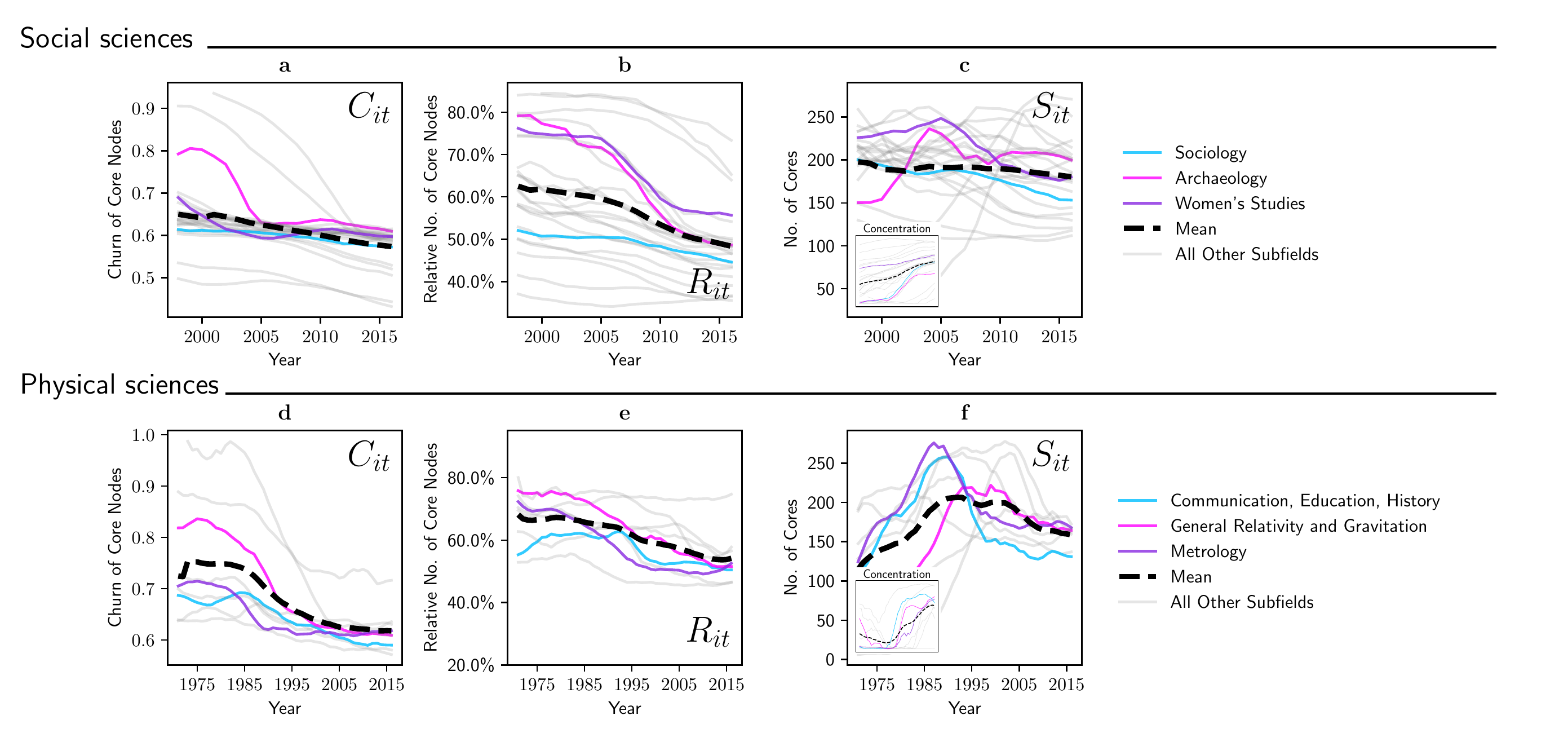}
\caption{\textbf{Conceptual structures over time for subfields.} The plot shows changes in the core/periphery organization for each of the individual subfields. The dashed black line represents the average trend across various subfields, while the other, thinner lines correspond to individual subfields. Trends over time appear consistent for most subfields. Notably, there is a discernible decrease in the churn of core concepts (a,d) and the relative size of the cores (b,c) in both the social and physical sciences. We also see a slight decline in the number of cores for the social sciences and an increasing number of cores followed by a flattening-out phase in recent years for the physical sciences. This growth is paired with a heightened concentration of concepts within a small number of core/periphery structures, as indicated by the Herfindahl index (refer to the inset plots on c and f).}
\label{fig:over_time_subfields}
\end{figure}
\pagebreak

\pagebreak

\begin{figure}[ht!]
\centering
\includegraphics[width=\textwidth]{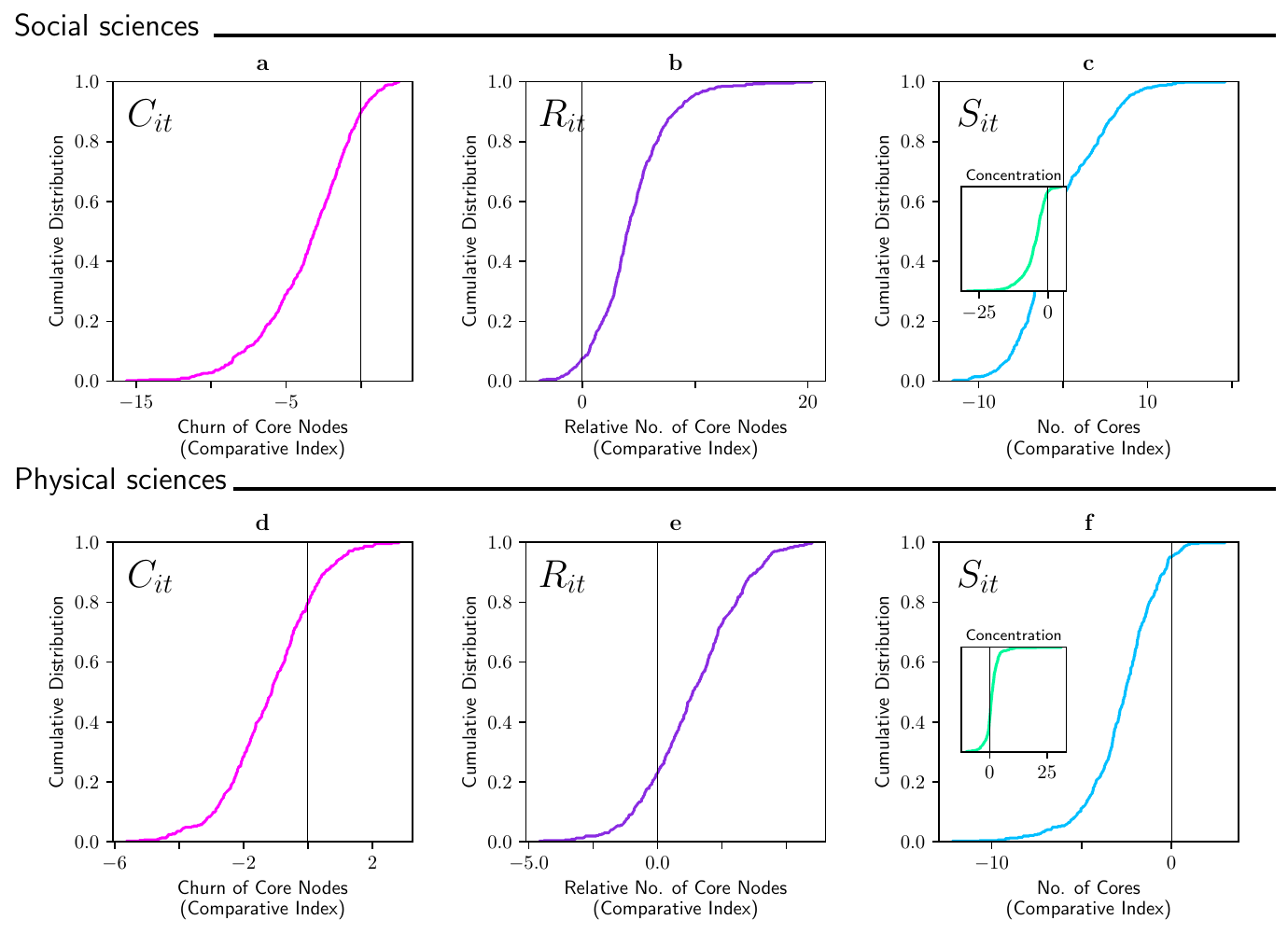}
\caption{\textbf{Comparative index of core/periphery properties against random networks.} This figure presents cumulative distributions that illustrate the deviation of core/periphery structures in observed networks from those in comparable randomized networks. For each field and year in our dataset, we maintained the original degree distribution by redistributing concepts across papers to create 50 random networks per observation. These networks preserve the number of appearances per concept and the number of concepts per paper. After projecting these bipartite networks into concept networks, we calculated core/periphery properties and derived a comparative index for each property by subtracting the mean measure of the random networks from the observed measure and normalizing by the standard deviation. The churn in core nodes for observed networks was consistently lower than in random networks across both social (top, left) and physical (bottom, left) sciences. Observed networks also exhibited a larger relative core size (top and bottom, center) compared to random ones. The number of cores in physical sciences observed networks was typically fewer than in their random counterparts, while in social sciences, the observed and random networks' number of cores were similar. Interestingly, the physical sciences showed higher concentration in observed networks compared to random ones, suggesting a more cumulative knowledge structure, whereas the social sciences displayed the opposite trend, potentially reflecting disciplinary differences in knowledge accumulation.}
\label{fig:random_networks}
\end{figure}

\pagebreak

\pagebreak

\begin{figure}[ht!]
\centering
\includegraphics[width=\textwidth]{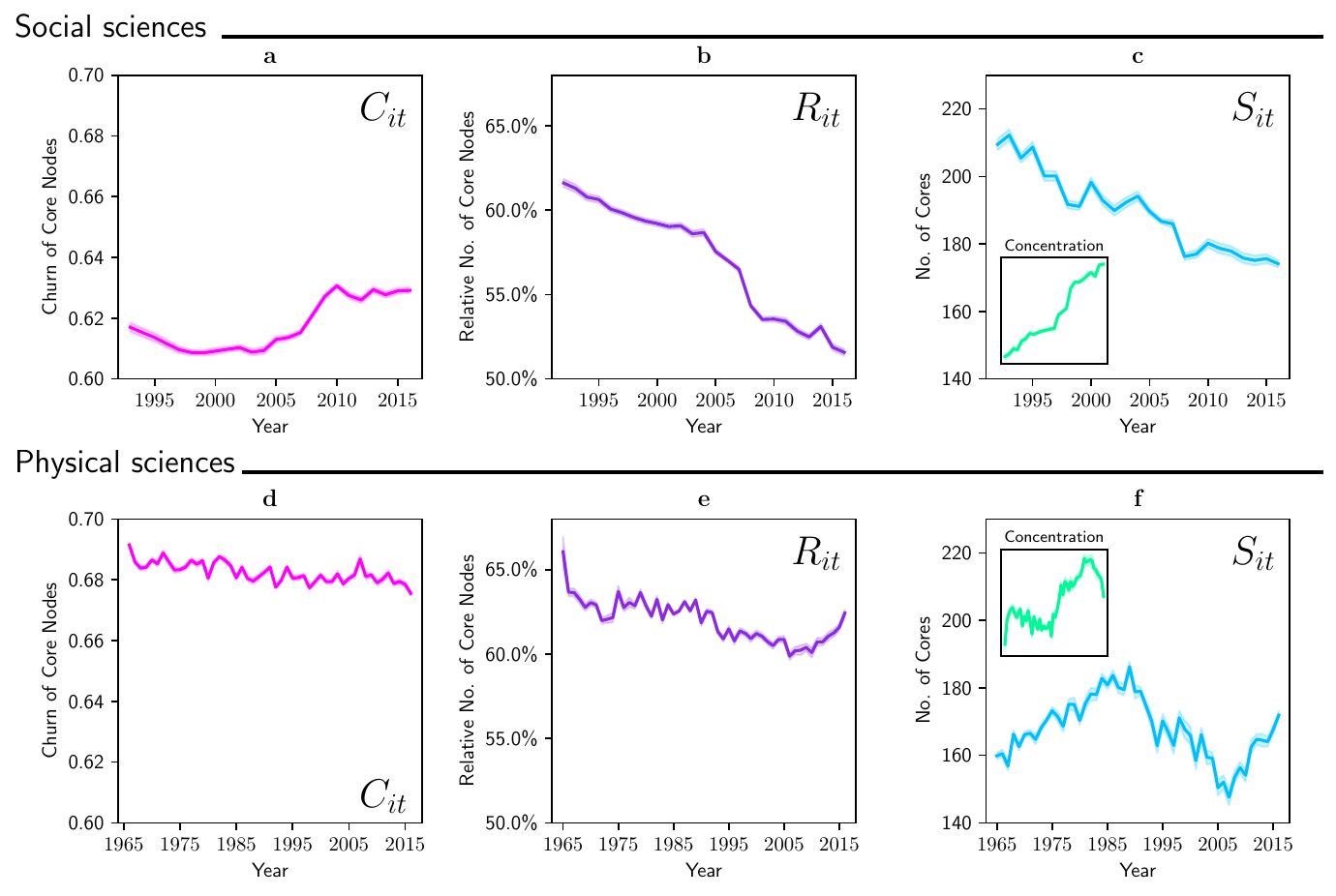}
\caption{\textbf{Conceptual structures over time after adjusting for expected properties in comparable random networks---within field null model.} These plots show annual changes in core/periphery organization after adjusting for the expected value of each core/periphery property in comparable random networks, using the within-field-and-year bipartite null model. In this model, we maintained the number of concepts per article while randomly reassigning the connections between articles and concepts within each field and year. This randomization generated comparable random networks, from which we then created concept-only networks by projecting the reshuffled bipartite structures. Adjustments were made using a regression-based approach, wherein the dependent variable was the core/periphery property of interest in the observed network and the values of the core/periphery property in random networks were included as a control. Each plot depicts the conditional yearly mean of a specific core/periphery property, derived from fixed effects regression models that also control for variations within subfields.}
\label{fig:R_adjusted_B2}
\end{figure}

\pagebreak

\pagebreak

\begin{figure}[ht!]
\centering
\includegraphics[width=\textwidth]{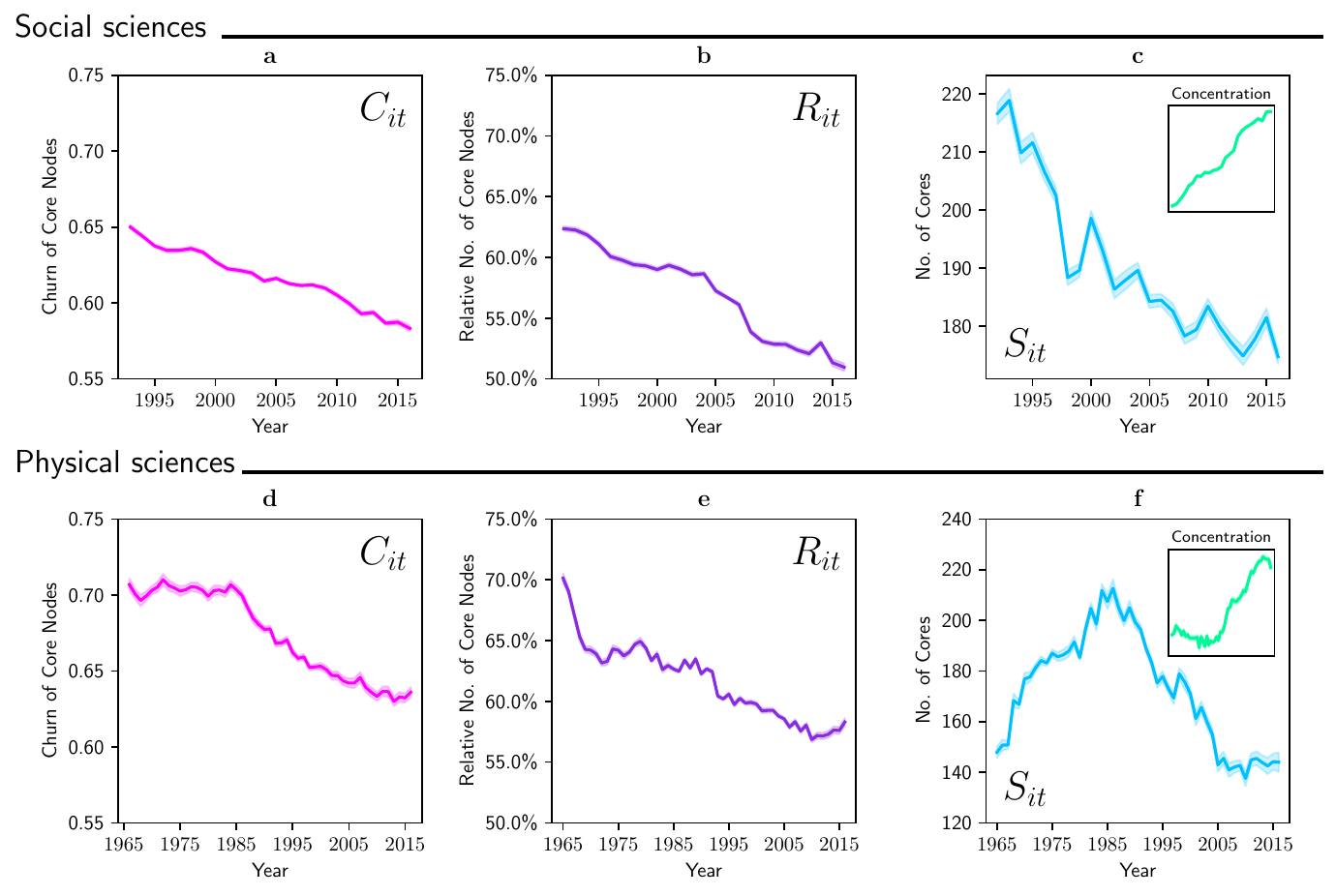}
\caption{\textbf{Conceptual structures over time after adjusting for expected properties in comparable random networks---across field null model.} These plots show annual changes in core/periphery organization after adjusting for the expected value of each core/periphery property in comparable random networks, using the across-field-and-within-year bipartite null model. In this model, we maintained the number of concepts per article while randomly reassigning the connections between articles and concepts across fields but within years. This randomization generated comparable random networks, from which we then created concept-only networks by projecting the reshuffled bipartite structures. Adjustments were made using a regression-based approach, wherein the dependent variable was the core/periphery property of interest in the observed network and the values of the core/periphery property in random networks were included as a control. Each plot depicts the conditional yearly mean of a specific core/periphery property, derived from fixed effects regression models that also control for variations within subfields.}
\label{fig:R_adjusted_FB2}
\end{figure}

\pagebreak

\pagebreak

\begin{figure}[ht!]
\centering
\includegraphics[width=\textwidth]{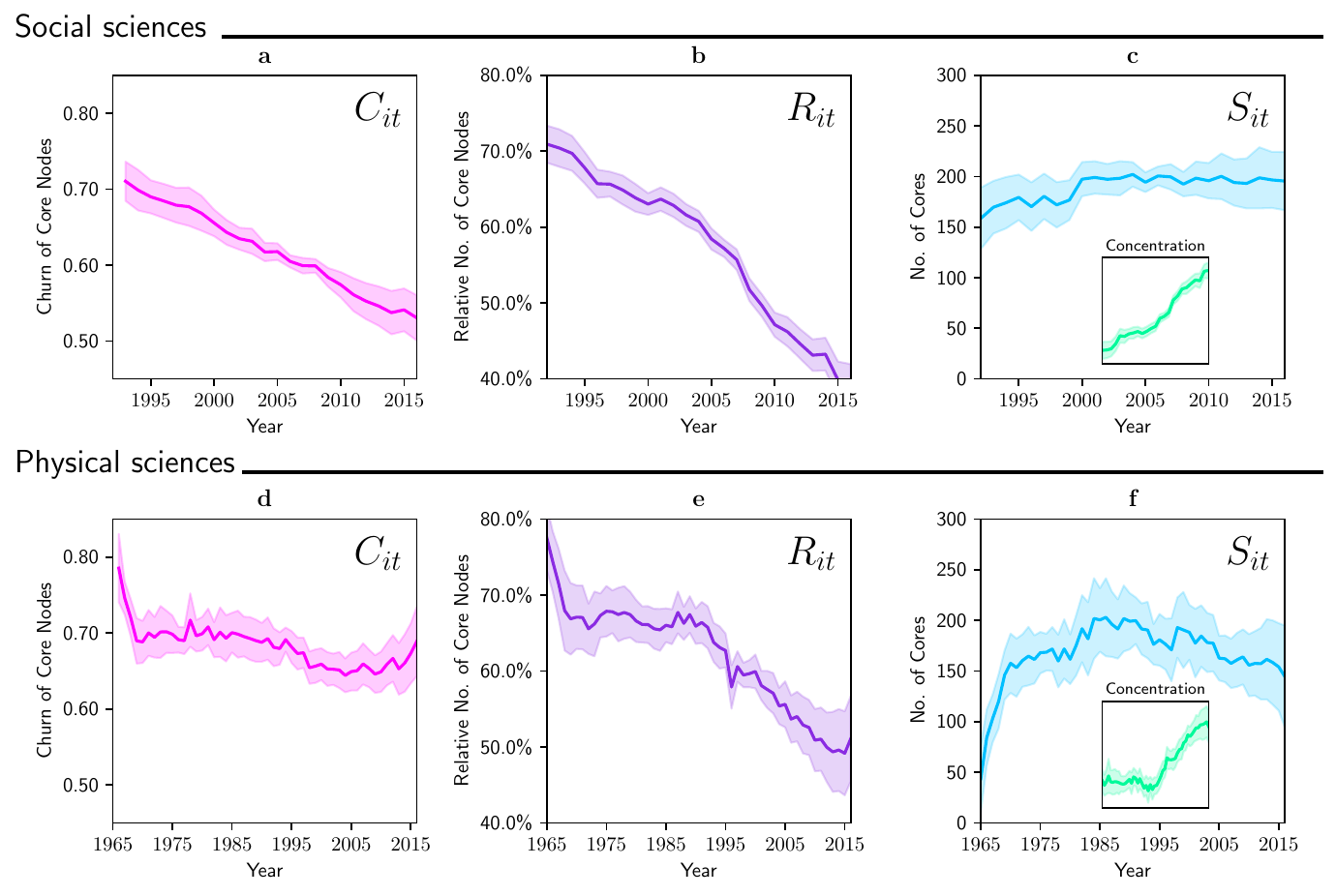}
\caption{\textbf{Conceptual structures over time after adjusting for linguistic and publication practices.} The plots show changes in the core/periphery organization after adjusting for changes in linguistic or publication practices over time. Adjustments were made using a regression-based approach, wherein the dependent variable was the core/periphery property of interest. Control variables included field $\times$ year level measures of the average \emph{Number of authors per paper}, \emph{Number of citations made per paper}, \emph{Length of sentences in paper abstracts}, \emph{Number of sentences in paper abstracts}, \emph{Length of words in paper abstracts}, \emph{Number of words in paper abstracts}, \emph{Fraction of nouns in paper abstracts}, \emph{Fraction of adjectives in paper abstracts}, \emph{Fraction of verbs in paper abstracts}, and \emph{Fraction of adverbs in paper abstracts}. The plot then shows the conditional mean of each core/periphery property by year. Models also included subfield fixed effects.}
\label{fig:language_controls_adjusted}
\end{figure}

\pagebreak

\begin{table}[ht!]
% \begin{center}
\centering
\begin{threeparttable}
\def\sym#1{\ifmmode^{#1}\else\(^{#1}\)\fi}
\caption{Regression models of over time trends}
\begin{tabular}{l c c c c c c}
\toprule
& \multicolumn{3}{c}{Social sciences} & \multicolumn{3}{c}{Physical sciences} \\\cmidrule(lr){2-4}\cmidrule(lr){5-7}
                                             &\multicolumn{1}{c}{\shortstack{\(C_{it}\): Churn of \\ core nodes }}&\multicolumn{1}{c}{\shortstack{\(R_{it}\): No. of \\ core nodes (\%)}}&\multicolumn{1}{c}{\shortstack{\(S_{it}\): No. of \\ cores (log)}}&\multicolumn{1}{c}{\shortstack{\(C_{it}\): Churn of \\ core nodes}}&\multicolumn{1}{c}{\shortstack{\(R_{it}\): No. of \\ core nodes (\%)}}&\multicolumn{1}{c}{\shortstack{\(S_{it}\): No. of \\ cores (log)}}\\
\midrule
Year       & $-0.0049^{***}$ & $-0.0087^{***}$ & $-0.5812^{*}$ & $-0.0042^{***}$ & $-0.0039^{***}$ & $2.5360^{***}$ \\
           & $(0.0002)$      & $(0.0002)$      & $(0.2301)$    & $(0.0001)$      & $(0.0001)$      & $(0.0864)$     \\
\hline
R$^2$      & $0.4452$        & $0.7088$        & $0.0110$      & $0.7076$        & $0.6958$        & $0.5471$       \\\hline
\multicolumn{7}{l}{}
\end{tabular}
% add table notes
\begin{tablenotes}

\item \emph{Notes:} Estimates are from ordinary-least-squares regressions. Models predict the core/periphery parameters as a function of year, separately for the social sciences and physical sciences. In all our models, we observed that the coefficient for \emph{Year} has a negative and statistically significant relationship with both the churn of core concepts and the relative size of the cores. \emph{Year} is negatively and significantly associated with the number of core/periphery structures in the social sciences, while it exhibits a positive and significant association with the number of such structures in the physical sciences. These results provide statistical support for the over time trends shown in the main manuscript. 
\item {*}p<0.05; {**}p<0.01; {***}p<0.001
\end{tablenotes}
\label{table:over_time_regressions}
% \end{center}
% \end{table}
\end{threeparttable}
\end{table}

\pagebreak

%%%%%%%%%%%%%%%%%%%%%%%%%%%%%%%%%%%%%%
%% BEGIN TABLE CONSENSUS REGRESSION %% 
%%%%%%%%%%%%%%%%%%%%%%%%%%%%%%%%%%%%%%

% format page as landscape
%\begin{landscape}

% remove page number
\thispagestyle{empty}
\pagestyle{empty}

{% begin table format

% set font size for table
\scriptsize

% squeeze columns together
\setlength{\tabcolsep}{2pt}

% squeeze rows together
\renewcommand{\arraystretch}{1}

% begin table header
\begin{table}[htbp]\centering
\begin{threeparttable}
\def\sym#1{\ifmmode^{#1}\else\(^{#1}\)\fi}
\caption{Regressions models of scientific consensus}
\label{table:ConsensusRegressions}
\begin{tabular}{l*{2}{c}}
\toprule

% add table data
\input{raw/table_consensus_regressions}\\
\bottomrule

% close tabular
\end{tabular}

% add table notes
\begin{tablenotes}

\item \emph{Notes:} Estimates are from ordinary-least-squares regressions. Models predict the use of consensus terms in physical science paper abstracts as a function of conceptual structure. Robust standard errors are shown in parentheses; p-values correspond to two-tailed tests. 
\item {*}p<0.05; {**}p<0.01; {***}p<0.001

\end{tablenotes}

% close table
\end{threeparttable}
\end{table}

% end table format
}%

% end landscape
%\end{landscape}

%%%%%%%%%%%%%%%%%%%%%%%%%%%%%%%%%%%%
%% END TABLE CONSENSUS REGRESSION %% 
%%%%%%%%%%%%%%%%%%%%%%%%%%%%%%%%%%%%

\pagebreak

%%%%%%%%%%%%%%%%%%%%%%%%%%%%%%%%%%%%%%
%% BEGIN TABLE MECHANISM VALIDATION %% 
%%%%%%%%%%%%%%%%%%%%%%%%%%%%%%%%%%%%%%

% format page as landscape
\begin{landscape}

{% begin table format

% set font size for table
\footnotesize

% remove page number
\thispagestyle{empty}

% squeeze columns together
% \setlength{\tabcolsep}{3pt}

% squeeze rows together
% \renewcommand{\arraystretch}{1}

% begin table header
\begin{table}[htbp]\centering
\begin{threeparttable}
\def\sym#1{\ifmmode^{#1}\else\(^{#1}\)\fi}
\caption{Properties of core/periphery concepts}
\label{table:MechanismValidation}
\begin{tabular}{lcccccccc}
\toprule

% add header
 &	\multicolumn{4}{c}{Social sciences}	&	\multicolumn{4}{c}{Physical sciences} \\ \cmidrule(lr){2-5}\cmidrule(lr){6-9}
 &	\multicolumn{1}{c}{\shortstack{Core \\ concepts}}&	\multicolumn{1}{c}{\shortstack{Periphery \\ concepts}}&&&	\multicolumn{1}{c}{\shortstack{Core \\ concepts}}	&	\multicolumn{1}{c}{\shortstack{Periphery \\ concepts}} \\ \cmidrule(lr){2-5}\cmidrule(lr){6-9}
		&	Mean/SD	&	Mean/SD	&	Welch's t	&	p-value	&	Mean/SD	&	Mean/SD	&	Welch's t	&	p-value	\\\cmidrule(lr){2-5}\cmidrule(lr){6-9}
Age of concept	&	21.3614	&	22.6192	&	-42.1343	&	<0.0001	&	47.8508	&	45.4913	&	5.5577	&	<0.0001	\\
	&	4.4163	&	2.8170	&		&		&	17.1380	&	16.1376	&		&		\\
Number of words in concept	&	1.4710	&	1.5531	&	-15.4247	&	<0.0001	&	1.7075	&	1.8158	&	9.1733	&	<0.0001	\\
	&	0.7065	&	0.5993	&		&		&	0.7089	&	0.7010	&		&		\\
Number of characters in concept	&	11.0221	&	11.8887	&	-19.7035	&	<0.0001	&	12.7720	&	13.6469	&	-9.8500	&	<0.0001	\\
	&	5.4234	&	5.3565	&		&		&	6.3366	&	6.4004	&		&		\\
Number of digits in concept	&	0.0641	&	0.0198	&	14.1870	&	<0.0001	&	0.0373	&	0.0510	&	-8.7761	&	<0.0001	\\
	&	0.4949	&	0.2433	&		&		&	0.2853	&	0.3406	&		&		\\
Number of subfields using concept	&	11.9393	&	10.1756	&	28.1218	&	<0.0001	&	4.2871	&	3.4499	&	-2.6978	&	0.0070	\\
	&	7.8256	&	7.5551	&		&		&	3.1953	&	2.7665	&		&		\\
Number of papers using concept	&	217.5179	&	196.4606	&	6.1072	&	<0.0001	&	157.0659	&	129.4299	&	18.4176	&	<0.0001	\\
	&	530.5200	&	294.9789	&		&		&	204.5506	&	98.3665	&		&		\\
Concept is Physics Stack Exchange tag	&		&		&		&		&	0.0662	&	0.0468	&	12.5400	&	<0.0001	\\
%N	&	31921	&	28330	&		&		&	12893	&	6004	&		&		\\
	\bottomrule

\end{tabular}

% add table notes
\begin{tablenotes}

\item \emph{Notes:} This table compares the properties of core/periphery concepts in the social and physical sciences. As we detail in subsequent analyses, many concepts move to and from the core/periphery (and back again) over time. For the purposes of this table, we defined core concepts as those that were observed more often in the core than the periphery (separately by field). All other concepts were defined as periphery concepts.

\end{tablenotes}

% close table
\end{threeparttable}
\end{table}

% end table format
}%

% end landscape
\end{landscape}

%%%%%%%%%%%%%%%%%%%%%%%%%%%%%%%%%%%%
%% END TABLE MECHANISM VALIDATION %% 
%%%%%%%%%%%%%%%%%%%%%%%%%%%%%%%%%%%%

%%%%%%%%%%%%%%%%%%%%%%%%%%%%%%%%%%%%%%%%%%%%
%% BEGIN TABLE RANDOM NETWORKS REGRESSION %% 
%%%%%%%%%%%%%%%%%%%%%%%%%%%%%%%%%%%%%%%%%%%%

% format page as landscape
%\begin{landscape}

% remove page number
\thispagestyle{empty}
\pagestyle{empty}

{% begin table format

% set font size for table
\scriptsize

% squeeze columns together
\setlength{\tabcolsep}{2pt}

% squeeze rows together
\renewcommand{\arraystretch}{1}

% begin table header
\begin{table}[htbp]\centering
\begin{threeparttable}
\def\sym#1{\ifmmode^{#1}\else\(^{#1}\)\fi}
\caption{Regression models of paper disruptiveness based on comparative core/periphery indexes}
\label{table:RandomNetworkRegressions}
\begin{tabular}{l*{2}{c}}
\toprule

% add table data
\input{raw/table_random_network_regressions_zscores_B2_rr2}\\

% add bottom rule
\bottomrule

% close tabular
\end{tabular}

% add table notes
\begin{tablenotes}

\item \emph{Notes:} Estimates are from ordinary-least-squares regressions. In these models, the core/periphery predictors are the comparative indexes that quantify deviations of the observed network properties from those in comparable random networks. To mitigate the effects of large comparative indexes, we apply an inverse hyperbolic sine transformation to the predictors. The models predict the disruptiveness of papers based on the comparative index of the conceptual structure of their field, with separate analyses for the social sciences (Model 1) and physical sciences (Model 2). Disruptiveness is measured using Funk and Owen-Smith's (\citeyearpar{funk2017}) CD index, based on citations made as of the year 2017. Wald tests reported below each model assess the significance of the core/periphery comparative indexes in improving model fit. Robust standard errors are in parentheses; p-values correspond to two-tailed tests.
\item {*}p<0.05; {**}p<0.01; {***}p<0.001

\end{tablenotes}

% close table
\end{threeparttable}
\end{table}

% end table format
}%

% end landscape
%\end{landscape}

%%%%%%%%%%%%%%%%%%%%%%%%%%%%%%%%%%%%%%%%%%
%% END TABLE RANDOM NETWORKS REGRESSION %% 
%%%%%%%%%%%%%%%%%%%%%%%%%%%%%%%%%%%%%%%%%%

%%%%%%%%%%%%%%%%%%%%%%%%%%%%%%%%%%%%%%%%%%%%%%%%%%%%%%%%%%%%%%%%%%
%% BEGIN TABLE OUTCOMES REGRESSION (RANDOM NETWORK ADJUSTED B2) %% 
%%%%%%%%%%%%%%%%%%%%%%%%%%%%%%%%%%%%%%%%%%%%%%%%%%%%%%%%%%%%%%%%%%

% remove page number
\thispagestyle{empty}
\pagestyle{empty}

{% begin table format

% set font size for table
\scriptsize

% squeeze columns together
\setlength{\tabcolsep}{2pt}

% squeeze rows together
\renewcommand{\arraystretch}{1}

% begin table header
\begin{table}[htbp]
\centering
\begin{threeparttable}
\def\sym#1{\ifmmode^{#1}\else\(^{#1}\)\fi}
\captionsetup{justification=centering}
\caption{Regression models of paper disruptiveness adjusted for core/periphery properties in comparable random networks}
\label{table:RandomNetworkRegressionsAdjustedB2}
\begin{tabular}{l*{4}{c}}
\toprule

% add table data
\input{raw/table_random_network_adjusted_regressions_B2_rr2}\\\bottomrule

% close tabular
\end{tabular}

% add table notes
\begin{tablenotes}

\item \emph{Notes:} Estimates are from ordinary-least-squares regressions. In these models, the core/periphery properties observed in the comparable random networks are included as controls. Note that the sample size is larger in our main models because there are repeated observations for each paper, one for each of the 50 random networks. For this analysis, the comparable random networks were derived by randomly shuffling the concepts associated with each article, within each field and year, while ensuring that each concept was associated with the same number of articles and each article was associated with the same number of concepts in both the observed and random networks. Subsequently, the randomly shuffled bipartite networks of papers $\times$ concepts was projected to yield a network of concepts. The models predict the disruptiveness of papers based on the conceptual structure of their field, with separate analyses for the social sciences (Model 1) and physical sciences (Model 2). Disruptiveness is measured using Funk and Owen-Smith's (\citeyearpar{funk2017}) CD index, based on citations made as of the year 2017. Wald tests reported below each model assess the significance of the observed core/periphery properties in improving model fit. Robust standard errors (clustered at the field $\times$ year level) are in parentheses; p-values correspond to two-tailed tests.
\item {*}p<0.05; {**}p<0.01; {***}p<0.001

\end{tablenotes}

% close table
\end{threeparttable}
\end{table}

% end table format
}%

%%%%%%%%%%%%%%%%%%%%%%%%%%%%%%%%%%%%%%%%%%%%%%%%%%%%%%%%%%%%%%%%
%% END TABLE OUTCOMES REGRESSION (RANDOM NETWORK ADJUSTED B2) %% 
%%%%%%%%%%%%%%%%%%%%%%%%%%%%%%%%%%%%%%%%%%%%%%%%%%%%%%%%%%%%%%%%

\end{document}

%% file: innovation_table.tex
%%%%%%%%%%%%%%%%%%%%%%%%%%%%%%%%%%%%%
%% BEGIN TABLE OUTCOMES REGRESSION %% 
%%%%%%%%%%%%%%%%%%%%%%%%%%%%%%%%%%%%%

% format page as landscape
%\begin{landscape}

% remove page number
% \thispagestyle{empty}
% \pagestyle{empty}

{% begin table format

% set font size for table
\scriptsize

% squeeze columns together
\setlength{\tabcolsep}{2pt}

% squeeze rows together
\renewcommand{\arraystretch}{1}

% begin table header
\begin{table}[htbp]
\centering
\caption{Regressions models of paper disruptiveness}
\begin{threeparttable}
\def\sym#1{\ifmmode^{#1}\else\(^{#1}\)\fi}

\label{table:OutcomesRegressions}
\begin{tabular}{l*{4}{c}}
\toprule

% add table data
\input{raw/table_outcomes_regressions}\\

% add bottom rule
\bottomrule

% close tabular
\end{tabular}

% add table notes
\begin{tablenotes}
\item \emph{Notes:} Estimates are from ordinary-least-squares regressions. Models predict the disruptiveness of papers as a function of the conceptual structure of their field, separately for the social sciences (Models 1-2) and physical sciences (Models 3-4). Disruptiveness is measured using Funk and Owen-Smith's (2017) $CD$ index, and is based on citations made as of the year 2017. Wald tests reported below each model evaluate whether the included core/periphery predictors significantly improve model fit. Robust standard errors are shown in parentheses; p-values correspond to two-tailed tests.
\item {*}p<0.05; {**}p<0.01; {***}p<0.001

\end{tablenotes}

% close table
\end{threeparttable}
\label{table:innovation_table}
\end{table}

% end table format
}%

% end landscape
%\end{landscape}

%%%%%%%%%%%%%%%%%%%%%%%%%%%%%%%%%%%
%% END TABLE OUTCOMES REGRESSION %% 
%%%%%%%%%%%%%%%%%%%%%%%%%%%%%%%%%%%

%% file: raw/table_outcomes_regressions.tex
 
                                             &\multicolumn{2}{c}{\shortstack{Social Sciences}}&\multicolumn{2}{c}{\shortstack{Physical Sciences}}\\\cmidrule(lr){2-3}\cmidrule(lr){4-5}
                                             &\multicolumn{1}{c}{(1)}   &\multicolumn{1}{c}{(2)}   &\multicolumn{1}{c}{(3)}   &\multicolumn{1}{c}{(4)}   \\
\midrule
$\textit{C}\textsubscript{it}$: Churn of core nodes&         0.2939***&         0.3335***&         0.1697***&         0.1729***\\
                                             &       (0.0135)   &       (0.0132)   &       (0.0189)   &       (0.0191)   \\
$\textit{R}\textsubscript{it}$: No. of core nodes (\%)&         0.0040***&         0.0047***&         0.0006***&         0.0007***\\
                                             &       (0.0001)   &       (0.0001)   &       (0.0001)   &       (0.0001)   \\
$\textit{S}\textsubscript{it}$: No. of cores &         0.0001***&         0.0002***&        -0.0000   &        -0.0000   \\
                                             &       (0.0000)   &       (0.0000)   &       (0.0000)   &       (0.0000)   \\
No. of authors                               &                  &        -0.0041***&                  &        -0.0010***\\
                                             &                  &       (0.0001)   &                  &       (0.0001)   \\
No. of references                            &                  &        -0.0026***&                  &        -0.0015***\\
                                             &                  &       (0.0000)   &                  &       (0.0000)   \\
Field fixed effects                          &            Yes   &            Yes   &            Yes   &            Yes   \\
Year fixed effects                           &            Yes   &            Yes   &            Yes   &            Yes   \\
\midrule
N                                            &        1765763   &        1765682   &         424282   &         418327   \\
Adjusted R2                                  &           0.06   &           0.11   &           0.02   &           0.03   \\
\midrule Wald tests for core/periphery predictors&                  &                  &                  &                  \\
F                                            &        1701.99   &        2534.13   &          72.61   &          81.92   \\
d.f.                                         &           3.00   &           3.00   &           3.00   &           3.00   \\
p-value                                      &           0.00   &           0.00   &           0.00   &           0.00   \\
 

%% file: raw/table_consensus_regressions.tex
 
                                             &\multicolumn{1}{c}{(1)}   &\multicolumn{1}{c}{(2)}   \\
\midrule
$\textit{C}\textsubscript{it}$: Churn of core nodes&        -1.4168** &        -1.3840** \\
                                             &       (0.4502)   &       (0.4582)   \\
$\textit{R}\textsubscript{it}$: No. of core nodes (\%)&         0.0025   &         0.0010   \\
                                             &       (0.0031)   &       (0.0031)   \\
$\textit{S}\textsubscript{it}$: No. of cores &        -0.0009** &        -0.0007*  \\
                                             &       (0.0004)   &       (0.0004)   \\
No. of nouns in abstract (log)               &        -0.1720***&        -0.1873***\\
                                             &       (0.0240)   &       (0.0244)   \\
No. of adjectives in abstract (log)          &         0.5254***&         0.5227***\\
                                             &       (0.0217)   &       (0.0222)   \\
No. of verbs in abstract (log)               &        -0.0260   &        -0.0194   \\
                                             &       (0.0266)   &       (0.0269)   \\
No. of adverbs in abstract (log)             &         0.1889***&         0.1908***\\
                                             &       (0.0140)   &       (0.0142)   \\
No. of authors                               &                  &         0.0024   \\
                                             &                  &       (0.0024)   \\
No. of references                            &                  &         0.0112***\\
                                             &                  &       (0.0010)   \\
Constant                                     &        -0.7697** &        -0.8276***\\
                                             &       (0.2375)   &       (0.2414)   \\\midrule
Field fixed effects                          &            Yes   &            Yes   \\
Year fixed effects                           &            Yes   &            Yes   \\
\midrule
N                                            &          80599   &          79669   \\
Mcfadden's pseudo-R2                         &           0.02   &           0.02   \\
 

%% file: raw/table_random_network_regressions_zscores_B2_rr2.tex
 
                                             &\multicolumn{1}{c}{\shortstack{Social Sciences}}&\multicolumn{1}{c}{\shortstack{Physical Sciences}}\\\cmidrule(lr){2-2}\cmidrule(lr){3-3}
                                             &\multicolumn{1}{c}{(1)}   &\multicolumn{1}{c}{(2)}   \\
\midrule
$\textit{C}\textsubscript{it}$: Churn of core nodes (Comparative Index)&         0.0034***&         0.0025***\\
                                             &       (0.0002)   &       (0.0004)   \\
$\textit{R}\textsubscript{it}$: No. of core nodes (\%) (Comparative Index)&         0.0014***&         0.0022***\\
                                             &       (0.0002)   &       (0.0003)   \\
$\textit{S}\textsubscript{it}$: No. of cores (Comparative Index)&         0.0058***&        -0.0010   \\
                                             &       (0.0002)   &       (0.0006)   \\
No. of authors                               &        -0.0040***&        -0.0009***\\
                                             &       (0.0001)   &       (0.0001)   \\
No. of references                            &        -0.0025***&        -0.0015***\\
                                             &       (0.0000)   &       (0.0000)   \\
Field fixed effects                          &            Yes   &            Yes   \\
Year fixed effects                           &            Yes   &            Yes   \\
\midrule
N                                            &        1765682   &         416294   \\
Adjusted R2                                  &           0.10   &           0.03   \\
\midrule Wald tests for core/periphery predictors&                  &                  \\
F                                            &         366.89   &          28.55   \\
d.f.                                         &           3.00   &           3.00   \\
p-value                                      &           0.00   &           0.00   \\
 

%% file: raw/table_random_network_adjusted_regressions_B2_rr2.tex
 
                                             &\multicolumn{1}{c}{\shortstack{Social Sciences}}&\multicolumn{1}{c}{\shortstack{Physical Sciences}}\\\cmidrule(lr){2-2}\cmidrule(lr){3-3}
                                             &\multicolumn{1}{c}{(1)}   &\multicolumn{1}{c}{(2)}   \\
\midrule
$\textit{C}\textsubscript{it}$: Churn of core nodes (Observed Network)&         0.2055*  &         0.1953***\\
                                             &       (0.0800)   &       (0.0424)   \\
$\textit{R}\textsubscript{it}$: No. of core nodes (\%) (Observed Network)&         0.0047***&         0.0009** \\
                                             &       (0.0005)   &       (0.0003)   \\
$\textit{S}\textsubscript{it}$: No. of cores (Observed Network)&        -0.0002*  &        -0.0000   \\
                                             &       (0.0001)   &       (0.0000)   \\
$\textit{C}\textsubscript{it}$: Churn of core nodes (Random Network)&         0.0606   &        -0.0099   \\
                                             &       (0.0851)   &       (0.0344)   \\
$\textit{R}\textsubscript{it}$: No. of core nodes (\%) (Random Network)&         0.0010   &        -0.0003   \\
                                             &       (0.0005)   &       (0.0003)   \\
$\textit{S}\textsubscript{it}$: No. of cores (Random Network)&         0.0005***&         0.0000   \\
                                             &       (0.0001)   &       (0.0000)   \\
No. of authors                               &        -0.0042***&        -0.0010***\\
                                             &       (0.0003)   &       (0.0001)   \\
No. of references                            &        -0.0026***&        -0.0015***\\
                                             &       (0.0003)   &       (0.0001)   \\
Field fixed effects                          &            Yes   &            Yes   \\
Year fixed effects                           &            Yes   &            Yes   \\
\midrule
N                                            &       8.83e+07   &       2.08e+07   \\
Adjusted R2                                  &           0.11   &           0.03   \\
\midrule Wald tests for core/periphery predictors&                  &                  \\
F                                            &          28.24   &          10.28   \\
d.f.                                         &           3.00   &           3.00   \\
p-value                                      &           0.00   &           0.00   \\
 